\documentclass[prc,notitlepage, twocolumn,
showpacs,floatfix,nofootinbib,preprintnumbers,superscriptaddress,amsmath,amssymb,aps,longbibliography]{revtex4-1}

\usepackage[utf8]{inputenc}
\usepackage{graphicx}\usepackage{epsfig}
\usepackage{amsmath,amssymb}
\usepackage{bm}
\usepackage{color}
\usepackage{float}
\usepackage{dcolumn} 
\usepackage{enumerate}
\usepackage{multirow}
\usepackage{threeparttable}
\usepackage{here}
\usepackage[bookmarks=true,colorlinks,%
            citecolor=blue,linkcolor=blue,anchorcolor=blue,filecolor=blue,urlcolor=blue,%
           ]{hyperref}

\begin{document}

\title{Radial and orbital decomposition of charge radii of Ca nuclei: \texorpdfstring{\\}{}
Comparative study of Skyrme and Fayans functionals}

\author{T. Inakura}
\email{inakura@gmail.com}
\affiliation{
Office of Institutional Research and Decision Support, Tokyo Institute of Technology, Meguro, Tokyo 152-8550, Japan
}\affiliation{
Laboratory for Zero-Carbon Energy, Institute of Innovative Research, Tokyo Institute of Technology, Meguro, Tokyo 152-8550, Japan
}

\author{N. Hinohara}
\email{hinohara@nucl.ph.tsukuba.ac.jp}
\affiliation{
 Center for Computational Sciences, University of Tsukuba, Tsukuba, Ibaraki 305-8577, Japan
}
\affiliation{
 Faculty of Pure and Applied Sciences, University of Tsukuba, Tsukuba, Ibaraki 305-8571, Japan
}
\affiliation{Facility for Rare Isotope Beams, 
Michigan State University, East Lansing, Michigan 48824, USA}

\author{H. Nakada}
\email{nakada@faculty.chiba-u.jp}
\affiliation{
Department of Physics, Graduate School of Science, Chiba University, Yayoi-cho 1-33, Inage, Chiba 263-8522, Japan}

\date{\today}

\begin{abstract}
  We investigate the charge and point-proton radii of the Ca nuclei in detail in the density functional theory framework.
  As the Fayans energy density functional provides characteristic $N$-dependence, successfully describing the parabolic behavior of the differential charge radii in $20\leq N\leq 28$, we pose our particular focus on its physics origin, by decomposing them into the radial and orbital contributions. The results are compared with those from the Skyrme plus usual pairing functional, which is taken as a representative of the functionals having normal pairing channels.
  We point out that, because the enhancement of the differential charge radii in $N<20$ with the Fayans functional, which is contradictory with the data, has the origin parallel to the parabolic behavior in $20\leq N\leq 28$, it is significant to describe both $N$ regions simultaneously.
\end{abstract}

\maketitle

\section{Introduction}

Radius is one of the fundamental properties representing nuclear structure.
It has been known that the radii of stable nuclei are proportional
to $A^{1/3}$ in the first approximation, where $A$ is the mass number,
linked to the saturation of the densities~\cite{Heyde2004}.
This feature is a result of the balance between the attraction among nucleons
at low density
and the repulsive effects that become dominant at high density.
The self-consistent mean-field calculations
or the density functional theory (DFT)
have been developed and are now recognized as one of the standard approaches
to the nuclear structure.
Since the nuclei are self-bound systems,
self-consistent calculations in the DFT
provide a suitable framework for studying nuclear radii.
An accurate description of radii is fundamental to the DFT,
because the density $\rho_t(\mathbf{r})$ ($t=\mathrm{n},\mathrm{p}$) is usually taken
as a principal variable~\cite{Nakada2023}
and the mean-square radius
is the lowest-order moment of $\rho_t(\mathbf{r})$,
\begin{equation}
  \langle r^2\rangle_t
  = \frac{1}{N_t} \int d^3r\,r^2\,\rho_t(\mathbf{r}),
\label{ms-radius}\end{equation}
where $N_t$ is the neutron or proton number (\textit{i.e.}, $N$ or $Z$).

The nuclear charge radii are measured by electric probes~\cite{ADNDT2013},
and their data are much less ambiguous than those obtained by hadronic probes.
Owing to the recent development
of the laser spectroscopy experiments~\cite{LaserSpectroscopy},
abundant data on the differential charge radii
have been accumulated via the isotope-shift measurements,
by which many intriguing results beyond the $A^{1/3}$ rule have been reported.
The kinks at magic neutron numbers are among them:
the kinks at $N=126$ near Pb isotopes~\cite{ADNDT1987},
at $N=82$ in the Sn isotopes~\cite{Gorges2019},
and at $N=28$ in the Ca and Ni isotopes~\cite{GarciaRuiz2016,Sommer2022}.
Arguments linking the kinks to properties of the nucleonic interaction
or the energy density functional (EDF) have been given.
The kinks are attributed to a property of the spin-orbit interaction
between nucleons in Refs.~\cite{Sharma1995,Reinhard1995},
which could be congenial to relativistic approaches~\cite{Sharma1993}.
Additional effects of density dependence in the spin-orbit interaction
connected to the three-nucleon interaction
have been discussed~\cite{M3Y-P6a,Nakada2015,Nakada2020}.
In contrast,
Fayans proposed an EDF with an extended form of the pairing functional~\cite{Fayans1998,Fayans2000},
which has been claimed to play significant roles in the kinks~\cite{Reinhard2017}.

Another interesting result of the charge radii has been known in $^{40\textrm{--}48}$Ca.
The charge radii of the even-$N$ nuclei in this region
vary parabolically as a function of $N$.
This behavior is hard to account for.
There has been a suggestion from the shell model
that the parabolic behavior may be a result of proton excitation
across $Z=20$~\cite{Caurier2001}.
A correlation of the charge radii and the quadrupole collectivity
has also been argued~\cite{Brown2022}.
On the other hand,
the Fayans EDF reproduces the parabolic behavior
in the self-consistent calculations that do not activate proton excitations across $Z=20$~\cite{Fayans2000,Reinhard2017}.
However, even-odd staggering of the charge radii tends to be too strong
when applying the Fayans EDF with parameters reproducing the parabola well,
and the behavior of the charge radii in the neutron-deficient region~\cite{Miller2019},
which was argued in connection to the $\ell s$-closed nature of $N=20$ and called \textit{anti-kink} in Ref.~\cite{Nakada2019},
seems incorrect.

We investigate the charge radii of the Ca isotopes in the framework of nuclear DFT.
As mentioned above, the Fayans EDF has distinguished characters,
being able to describe the parabolic behavior of the charge radii in $20\leq N\leq 28$.
In Ref.~\cite{Reinhard2017},
this property has been pointed out to originate
in its specific pairing channel.
Despite the remarkable result,
it is not obvious whether this picture is truly reasonable.
We inspect the radial and orbital contributions to the radii
in comparison with other EDFs,
particularly a Skyrme EDF.
The results may help to discriminate the EDF in future experiments.


\section{Method}

\subsection{Skyrme and Fayans EDF}

The Skyrme and Fayans EDFs are the functionals of several sets of local densities, commonly written in the form:
\begin{align}
 E &= \int d^3r\,{\cal E}(\mathbf{r}); \\
 {\cal E}(\mathbf{r}) &= 
\sum_{t=\mathrm{n},\mathrm{p}}\frac{\tau_t(\mathbf{r})}{2M} +
 {\cal E}_{\rm ph}(\mathbf{r}) + {\cal E}_{\rm Coul}(\mathbf{r}) + {\cal E}_{\rm pair}(\mathbf{r}),
\end{align}
where ${\cal E}$ is composed of the kinetic energy term with the averaged nucleon mass $M$, nuclear particle-hole part, Coulomb, and pairing part.
In this paper, we assume the time-reversal symmetry, restricting ourselves to even-even nuclei.
For a Skyrme EDF, 
the time-even particle-hole part of  ${\cal E}$ is given by
\begin{align}
{\cal E}^{\rm Sk}_{\rm ph}[\rho,\tau, \mathbf{J}] &=  
{\cal E}^{\rm Sk}_{\rm v}[\rho,\tau]
+ {\cal E}^{\rm Sk}_{\rm s}[\rho]
+ {\cal E}^{\rm Sk}_{\rm ls}[\rho, \mathbf{J}],
\\
{\cal E}^{\rm Sk}_{\rm v}[\rho,\tau]
&=
\sum_{k=0,1} 
C_k^\rho[\rho_0] [\rho_k(\mathbf{r})]^2 + C_k^\tau
\rho_k(\mathbf{r})\tau_k(\mathbf{r}), \\
{\cal E}^{\rm Sk}_{\rm s}[\rho] &=
\sum_{k=0,1} 
C_k^{\Delta\rho} \rho_k(\mathbf{r})\Delta\rho_k(\mathbf{r}), \\
{\cal E}^{\rm Sk}_{\rm ls}[\rho, \mathbf{J}]
&= \sum_{k=0,1} 
C_k^{\nabla J} \rho_k(\mathbf{r}) \mathbf{\nabla}\cdot\mathbf{J}_k(\mathbf{r}),
\end{align}
where 
$\rho_k(\mathbf{r})$, $\tau_k(\mathbf{r})$, and $\mathbf{J}_k(\mathbf{r})$ are the local particle densities, kinetic densities, and spin-orbit densities \cite{Bender2003}, with $k=0$ and $1$ corresponding to the isoscalar and isovector parts; \textit{e.g.}, $\rho_0(\mathbf{r})=\rho_{\mathrm{n}}(\mathbf{r})+\rho_{\mathrm{p}}(\mathbf{r})$ and $\rho_1(\mathbf{r})=\rho_{\mathrm{n}}(\mathbf{r})-\rho_{\mathrm{p}}(\mathbf{r})$.
The $\rho_0$-dependent coupling constant is given in the form $C^\rho_k[\rho_0]
= C_k^\rho[0] + C_{kD}^{\rho} [\rho_0(\mathbf{r})]^\alpha$.

The particle-hole part of the Fayans EDF \cite{Fayans1998}, which does not contain $\tau_k(\mathbf{r})$,
is composed of the volume, surface, and spin-orbit parts,
\begin{align}
    {\cal E}^{\rm Fy}_{\rm ph}[\rho,\mathbf{J}] &= {\cal E}_{\rm v}^{\rm Fy}[\rho] + {\cal E}_{\rm s}^{\rm Fy}[\rho]
    + {\cal E}_{\rm ls}^{\rm Fy}[\rho,\mathbf{J}], \\
{\cal E}^{\rm Fy}_{\rm v}[\rho] &= \frac{\epsilon_F\rho_{\rm sat}}{3}
\left\{
a_+^{\rm v} \frac{1 - h_{1+}^{\rm v}[x_0(\mathbf{r})]^\sigma}{1 + h_{2+}^{\rm v} [x_0(\mathbf{r})]^\sigma} [x_0(\mathbf{r})]^2
\right. \nonumber \\ &\quad  \left.
+
a_-^{\rm v} \frac{1 - h_{1-}^{\rm v}x_0(\mathbf{r})}{1 + h_{2-}^{\rm v} x_0(\mathbf{r})}[x_1(\mathbf{r})]^2
\right\}
, \\
{\cal E}^{\rm Fy}_{\rm s}[\rho] &= \frac{\epsilon_F \rho_{\rm sat}}{3}
\frac{ a_+^{\rm s} r_s^2 [\nabla x_0(\mathbf{r})]^2}
{1 + h_+^{\rm s} [x_0(\mathbf{r})]^\sigma + h^{\rm s}_{\nabla} r_s^2 [\nabla x_0(\mathbf{r})]^2},\\
{\cal E}_{\rm ls}^{\rm Fy}[\rho,\mathbf{J}] &= 
\frac{4\epsilon_Fr_s^2}{3\rho_{\rm sat}}
[ \kappa \rho_0(\mathbf{r}) \mathbf{\nabla}\cdot \mathbf{J}_0(\mathbf{r})
+ \kappa'\rho_1(\mathbf{r}) \mathbf{\nabla}\cdot \mathbf{J}_1(\mathbf{r})].
\end{align}
The volume and surface terms of the Fayans EDF are given in a form analogous to the Pad\'{e} approximant. The density $\rho_k(\mathbf{r})$ is used in terms of $x_k(\mathbf{r})=\rho_k(\mathbf{r})/\rho_{\rm sat}$, normalized with the saturation density $\rho_{\rm sat}=0.16$\,fm$^{-3}$.
The parameters $\epsilon_F$ and $r_s$ are the Fermi energy and the Wigner-Seitz radius, whose values are given in Appendix~\ref{sec:FaNDF0}.
The volume term corresponds to the Skyrme coupling constants as
\begin{align}
    C_0^\rho[\rho_0] &= \frac{\epsilon_F a_+^{\rm v}}{3\rho_{\rm sat}}
     \frac{1 - h_{1+}^{\rm v}[x_0(\mathbf{r})]^\sigma}{1 + h_{2+}^{\rm v} [x_0(\mathbf{r})]^\sigma},\\
    C_1^\rho[\rho_0] &= \frac{\epsilon_F a_-^{\rm v}}{3\rho_{\rm sat}} 
     \frac{1 - h_{1-}^{\rm v}x_0(\mathbf{r})}{1 + h_{2-}^{\rm v} x_0(\mathbf{r})},
\end{align}
and the surface term of the Fayans EDF roughly corresponds to 
the isoscalar surface term $C_0^{\Delta\rho}\rho_0\Delta\rho_0$ of the Skyrme EDF with a density-dependent coupling constant.
The isovector coupling term $C_1^{\Delta\rho}$ is not present in the 
Fayans surface term, and the effective mass of the Fayans EDF is kept unity.
The spin-orbit term of the Fayans EDF has the same functional form as the extended Skyrme EDF~\cite{Reinhard1995}.
The two parameters $\kappa$ and $\kappa'$ are used that are related to the Skyrme coupling constants as
\begin{align}
 C_0^{\nabla J} &= \frac{4\epsilon_F r_s^2}{3\rho_{\rm sat}}\kappa, \\
 C_1^{\nabla J} &= \frac{4\epsilon_F r_s^2}{3\rho_{\rm sat}}\kappa'.
\end{align}
The Coulomb EDF is written as a sum of the direct and exchange contributions,
\begin{align}
{\cal E}_{\rm Coul}[\rho] &=  \frac{e^2}{2} \rho_{\mathrm{p}}(\mathbf{r}) 
\int d^3r'  \frac{\rho_{\mathrm{p}}(\mathbf{r}')}{|\mathbf{r}-\mathbf{r}'|}
\nonumber \\ &\quad 
 - \frac{3}{4} \left(\frac{3}{\pi}\right)^{\frac{1}{3}}
 e^2 [\rho_{\mathrm{p}}(\mathbf{r})]^{\frac{4}{3}}\left\{ 1 - h_{\rm Coul}[x_0(\mathbf{r})]^\sigma\right\},
\end{align}
where the point-proton density $\rho_{\mathrm{p}}(\mathbf{r})$ is employed instead of the charge density.
The Coulomb exchange energy is evaluated with the Slater approximation, apart from the Coulomb-nuclear correlation term controlled by the parameter $h_{\rm Coul}$.
In the Skyrme EDF, $h_{\rm Coul}=0$.

The pairing functionals of the Skyrme and Fayans EDFs are 
\begin{align}
\mathcal{E}_\mathrm{pair}^{\rm Sk}[\rho,\tilde{\rho},\tilde{\rho}^\ast]
&= \sum_{t={\rm n,p}} \frac{V_t}{4}\left[ 1 - \eta \frac{\rho_0(\mathbf{r})}{\rho_{\rm pair}}
\right] |\tilde{\rho}_t(\mathbf{r})|^2,
\label{mixed.pairing} \\
\mathcal{E}_\mathrm{pair}^{\rm Fy}[\rho,\tilde{\rho},\tilde{\rho}^\ast]
&= 
\frac{2\epsilon_F}{3\rho_{\rm sat}} 
\left\{ f^\xi_{\rm ex} 
+ h_+^\xi [x_{\rm pair}
(\mathbf{r})]^\gamma  
\right.\nonumber \\ &\quad \left.
+  h_\nabla^\xi r_s^2 [\nabla x_{\rm pair}(\mathbf{r})]^2
\right\} 
\sum_{t={\rm n,p}} |\tilde{\rho}_t(\mathbf{r})|^2,
\label{FayansEDF.pairing}
\end{align}
with $x_{\rm pair}(\mathbf{r})=\rho_0(\mathbf{r})/\rho_{\rm pair}$.
It is customary to set $\rho_{\rm pair}=\rho_{\rm sat}$.
In the Skyrme EDF, we use the mixed-type pairing with $\eta=1/2$
unless explicitly specified.
When $\gamma=1$, the Fayans pairing parameters $f^\xi_{\rm ex}$ and $h^\xi_+$ are related to the Skyrme pairing parameters as
\begin{align}
V_t &= \frac{8\epsilon_F}{3\rho_{\rm sat}} f^\xi_{\rm ex}, \\
\eta &= -\frac{h^\xi_+}{ f^\xi_{\rm ex}}.
\end{align}
On the other hand, the $h^\xi_\nabla$ term that depends on the derivative of the isoscalar density is not present in the standard density-dependent pairing used with the Skyrme EDF. 

The variational condition of the EDF leads to the 
Hartree-Fock-Bogoliubov equation,
\begin{align}
    \begin{pmatrix} 
    h_t-\lambda_t & \tilde{h}_t \\ -\tilde{h}_t^\ast & -h_t^\ast+\lambda_t
    \end{pmatrix}
\begin{pmatrix} \mathcal{U}_{\mu,t} \\ \mathcal{V}_{\mu,t}
\end{pmatrix}
= E_{\mu,t}
\begin{pmatrix} \mathcal{U}_{\mu,t} \\ \mathcal{V}_{\mu,t} \end{pmatrix}.
\end{align}
The particle-hole and pairing parts of the local mean-field Hamiltonian are given by
\begin{align}
    h_t(\mathbf{r};s',s) &=
    -\frac{\hbar^2}{2M}\nabla^2 \delta_{s's}
    + U_t(\mathbf{r})\delta_{s's}
\nonumber \\ & \quad
    + \frac{1}{i} (\mathbf{\nabla}\times \hat{\bm{\sigma}}_{s's})W_t(\mathbf{r})\cdot \mathbf{\nabla}, \\
    \tilde{h}_t(\mathbf{r};s',s) &=
    \tilde{U}_t(\mathbf{r}) \delta_{s's},
\end{align}
where $U_t(\mathbf{r}), 
W_t(\mathbf{r})$,
and $\tilde{U}_t(\mathbf{r})$ are
the central, 
spin-orbit, and pairing potentials, respectively,
$\lambda_t$ is the chemical potential,
and $\hat{\bm{\sigma}}$ stands for the Pauli matrix.
The potentials are calculated by the functional derivatives with respect to the corresponding densities
\begin{align}
U_t(\mathbf{r}) &= \frac{ \delta {\cal E}[\rho, \mathbf{J},\tilde{\rho},\tilde{\rho}^\ast]}
{\delta \rho_t}, \\
W_t(\mathbf{r}) &= \frac{ \delta {\cal E}[\rho, \mathbf{J},\tilde{\rho},\tilde{\rho}^\ast]}
{\delta (\mathbf{\nabla}\cdot\mathbf{J}_t)}, \\
\tilde{U}_t(\mathbf{r}) &= \frac{\delta {\cal E}[\rho, \mathbf{J},\tilde{\rho},\tilde{\rho}^\ast]}{\delta \tilde{\rho}^\ast_t}.
\end{align}
The pair gap is defined as a density-averaged pair potential,
\begin{align}
\Delta_t = \frac{1}{N_t}\int d^3 r \tilde{U}_t(\mathbf{r}) \rho_t(\mathbf{r}).
\end{align}

Because it plays an important role in the charge radius in the Ca isotope,
we explicitly write the rearrangement term,
the contribution to the central potential $U_t(\mathbf{r})$ from the pairing EDF~\cite{Reinhard2017},
\begin{align}
\frac{\delta{\cal E}_{\rm pair}^{\rm Sk}
}{\delta \rho_t}
&= - \frac{\eta}{\rho_{\rm pair}}
\sum_{t'={\rm n,p}}
\frac{V_{t'}}{4}|\tilde{\rho}_{t'}(\mathbf{r})|^2, \\
\frac{\delta{\cal E}_{\rm pair}^{\rm Fy}
}{\delta \rho_t}
&= \frac{2\epsilon_F}{3\rho_{\rm sat}} \left\{
\left[
\frac{\gamma h_+^\xi}{\rho_{\rm pair}} [x_{\rm pair}(\mathbf{r})]^{\gamma-1}
 -\frac{2h^\xi_\nabla r_s^2}{\rho_{\rm pair}^2}
 \Delta\rho_0 (\mathbf{r})\right] \right.
 \nonumber \\ &\quad \times
\sum_{t'={\rm n,p}}
|\tilde{\rho}_{t'}(\mathbf{r})|^2
\nonumber \\ &\quad  \left.
-
\frac{2h^\xi_\nabla r_s^2}{\rho_{\rm pair}^2}
\sum_{t'={\rm n,p}} 
{\rm Re}\left[
\tilde{\rho}_{t'}^\ast(\mathbf{r}) 
\mathbf{\nabla}\rho_0(\mathbf{r})\cdot
\mathbf{\nabla}\tilde{\rho}_{t'}(\mathbf{r}) \right] \right\}.
\label{eq:rearrangement}
\end{align}
The other terms of the potentials in the Fayans functionals are given in Appendix~\ref{sec:potential}.

\subsection{Differential radii and densities}

The mean-square radius of point protons or neutrons is related to their density distributions via Eq.~\eqref{ms-radius}.
In the present work, the spherical symmetry is assumed for all nuclei under investigation, and the densities are functions of $r=|\mathbf{r}|$.
While the charge radius and the charge density are primarily subject to the distribution of the point protons,
the finite-size and the relativistic corrections~\cite{Friar-Negele_1975,Kurasawa2019} are taken into account as summarized in Appendix~\ref{sec:chargedensity}.

The isotopic differences of the mean-square proton and charge radii are denoted by
\begin{align}
  \delta\langle r^2\rangle^{A_0,A}_\mathrm{p} &= \langle r^2\rangle_\mathrm{p}(^{A}\mathrm{Ca}) -  \langle r^2\rangle_\mathrm{p}(^{A_0}\mathrm{Ca}),
    \label{eq:diff_rp}\\
  \delta\langle r^2\rangle^{A_0,A}_\mathrm{ch} &= \langle r^2\rangle_\mathrm{ch}(^{A}\mathrm{Ca}) -  \langle r^2\rangle_\mathrm{ch}(^{A_0}\mathrm{Ca}),
  \label{eq:diff_rch}
\end{align}
with a reference nucleus $^{A_0}$Ca.
We also define the differential proton and charge densities as
\begin{align}
  \bar{\Delta}\rho_{\rm p}^{A_0,A}(r)&:= (4\pi r^2) r^2 \left[\rho_\mathrm{p}(r,\,^{A}\mathrm{Ca}) - \rho_\mathrm{p}(r,\,^{A_0}{\rm Ca})\right],
  \label{eq:diff_rhop}\\
\bar{\Delta}\rho_{\rm ch}^{A_0,A}(r)&:= (4\pi r^2) r^2 \left[\rho_\mathrm{ch}(r,\,^{A}\mathrm{Ca}) - \rho_\mathrm{ch}(r,\,^{A_0}{\rm Ca})\right],
\label{eq:diff_rhoch}
\end{align}
so that the integration of $\bar{\Delta}\rho_{\rm p}^{A_0,A}(r)$ [$\bar{\Delta}\rho_{\rm ch}^{A_0,A}(r)$] over $r$ should be equal to $\delta\langle r^2\rangle^{A_0,A}_\mathrm{p}$ [$\delta\langle r^2\rangle^{A_0,A}_\mathrm{ch}$].

In Sec.~\ref{subsec:orbital}, we will further consider the contributions of individual proton orbits $j$ to the proton radius and density.
For this purpose, we consider the following quantities:
\begin{align}
  \bar{\Delta}\rho_{{\rm p},j}^{A_0,A}(r)&:= (4\pi r^2) r^2 \sum_m \left[\big|\varphi_{\mathrm{p},jm}(\mathbf{r},\,^{A}\mathrm{Ca})\big|^2\right. \nonumber\\
    &\qquad\quad\left. - \big|\varphi_{\mathrm{p},jm}(\mathbf{r},\,^{A_0}\mathrm{Ca})\big|^2\right],
  \label{eq:diff_rhopj}
\end{align}
where $\varphi_{\mathrm{p},jm}$ is the wave functions of the proton single-particle (s.p.) orbit $j$ with the magnetic quantum number $m$.
We here use the quantum number $j$ also for representing the s.p. orbit, for simplicity.
The sum of $\bar{\Delta}\rho_{{\rm p},j}^{A_0,A}(r)$ over the occupied orbit $j$ is equal to $\bar{\Delta}\rho_{\rm p}^{A_0,A}(r)$ in Eq.~\eqref{eq:diff_rhop}.
For the radii, we consider the difference of the root-mean-square radii of the orbit,
\begin{align}
  \delta\big[\sqrt{\langle r^2\rangle}\big]^{A_0,A}_{\mathrm{p},j} &:= \bigg\{\int_0^\infty dr\,( 4\pi r^2) r^2 \nonumber\\
    &\qquad \times \frac{1}{2j+1}\sum_m \big|\varphi_{\mathrm{p},jm}(\mathbf{r},\,^{A}\mathrm{Ca})\big|^2 \bigg\}^{1/2} \nonumber\\
      &\quad - \bigg\{\int_0^\infty dr\,( 4\pi r^2) r^2 \nonumber\\
    &\qquad \times \frac{1}{2j+1}\sum_m \big|\varphi_{\mathrm{p},jm}(\mathbf{r},\,^{A_0}\mathrm{Ca})\big|^2 \bigg\}^{1/2}.
    \label{eq:diff_rpj}
\end{align}

\subsection{Numerical settings}

The coordinate-basis spherical code {\sc hfbrad}~\cite{HFBRAD} is employed to analyze the property of the charge radius in Ca isotopes.
We have also implemented the Fayans EDF in {\sc hfbrad}.
We set the box size $R_{\rm max} = 15$ fm, with the lattice size $\Delta r= 0.1 $ fm. We use the pairing cutoff of 15~MeV for Fayans EDF and 60 MeV for Skyrme EDFs.
The pairing strength for the mixed-type pairing of the Skyrme EDF is 
$V_{\rm n}=-283.7$ MeV fm$^3$, 
which has been adjusted to the neutron gap of 1.245 MeV in $^{120}$Sn~\cite{HFBRAD}.

\section{Results \label{sec:results}}

\subsection{Charge radii and charge densities}

\begin{figure}[H]
\begin{center}
\includegraphics[width=0.450\textwidth,keepaspectratio]{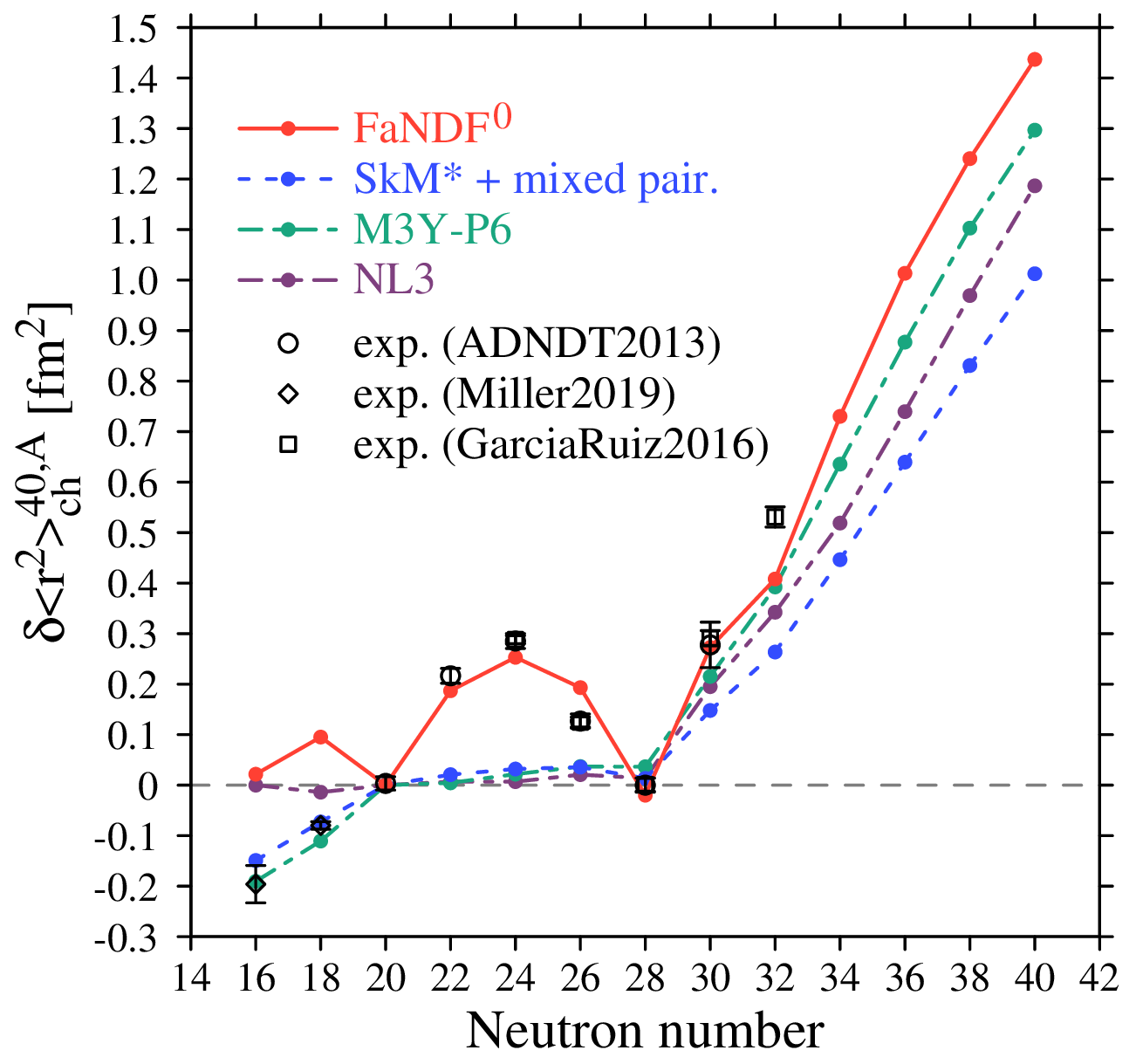}
\caption{Isotopic difference of charge radii of even-$N$ Ca isotopes, $\delta\langle r^2 \rangle^{40,A}_\mathrm{ch}$ calculated with FaNDF$^0$, SkM$^\ast$ and M3Y-P6 are compared with the experimental data~\cite{ADNDT2013,Miller2019,GarciaRuiz2016}. The charge radii calculated with NL3 are taken from Ref.~\cite{Lalazissis1999} for comparison.
}
\label{isotope.shift}
\end{center}
\end{figure}

We have calculated the charge radii of even-$N$ Ca isotopes with FaNDF$^0$ of the Fayans EDF~\cite{Fayans1998} and several other EDFs and interactions; Skyrme SkM$^\ast$~\cite{SkMs}, SLy4~\cite{SLy4}, UNEDF1~\cite{UNEDF1} and M3Y-type M3Y-P6~\cite{M3Y-P6,Nakada2014}.
For comparison, the result of NL3, which is a relativistic EDF, is also quoted from Ref.~\cite{Lalazissis1999}.
Figure~\ref{isotope.shift} shows the calculated differential charge radii $\delta\langle r^2\rangle^{40,A}_\mathrm{ch}$ of even-$N$ Ca isotopes 
and compares them with the experimental data~\cite{ADNDT2013,Miller2019,GarciaRuiz2016}.
No proton excitations across the $Z=20$ core take place in any calculations, and the $N$-dependence is ascribed to the influence of neutrons on the proton s.p. orbitals.
Only FaNDF$^0$ reproduces the parabola structure of charge radii in $^{40\textrm{--}48}$Ca as already discussed in Ref.~\cite{Reinhard2017}, while the other interactions yield almost constant charge radii in $^{40\textrm{--}48}$Ca.
In the neutron-deficient region, FaNDF$^0$ does not reproduce the charge radii in $N=16$ and 18;
the charge radii computed with the FaNDF$^0$ in $^{36,38}$Ca are larger than at $^{40}$Ca, while the experimental values decrease for decreasing $N$.
SkM$^\ast$ and M3Y-P6 reproduce the experimental data in $^{36,38}$Ca fairly well.
NL3 produces almost the constant charge radii in $^{36\textrm{--}48}$Ca.
Consequently, no EDFs successfully reproduce the charge radii in the whole region of $^{36\textrm{--}48}$Ca.
As mentioned in Introduction, the influence of the proton excitation across $Z=20$ including the shape fluctuation, which might be beyond the mean-field description, has been suggested for $^{42\textrm{--}46}$Ca in some studies~\cite{Caurier2001,Brown2022}.

For neutron-rich nuclei with $N>28$, all interactions reproduce and predict the rapid growth of the charge radii with the neutron number, 
though the growth rate depends on the EDFs.

Because the Fayans EDF gives characteristic behavior of the charge radii in $^{40\textrm{--}48}$Ca, we shall investigate in detail what gives rise to it.
In the following, we take SkM$^\ast$ as a representative of the usual EDFs and compare its results with those of the Fayans EDF.
While the Fayans-EDF parameters are further tuned in the functional Fy$(\Delta r)$~\cite{Reinhard2017} so as to reproduce the parabola structure of charge radii,
we use the original parameter FaNDF$^0$ in this paper,
whose result is close to that of Fy$(\Delta r)$.

\subsection{Charge and point-proton density distributions}

\begin{figure}[h]
\begin{center}
\includegraphics[width=0.450\textwidth,keepaspectratio]
{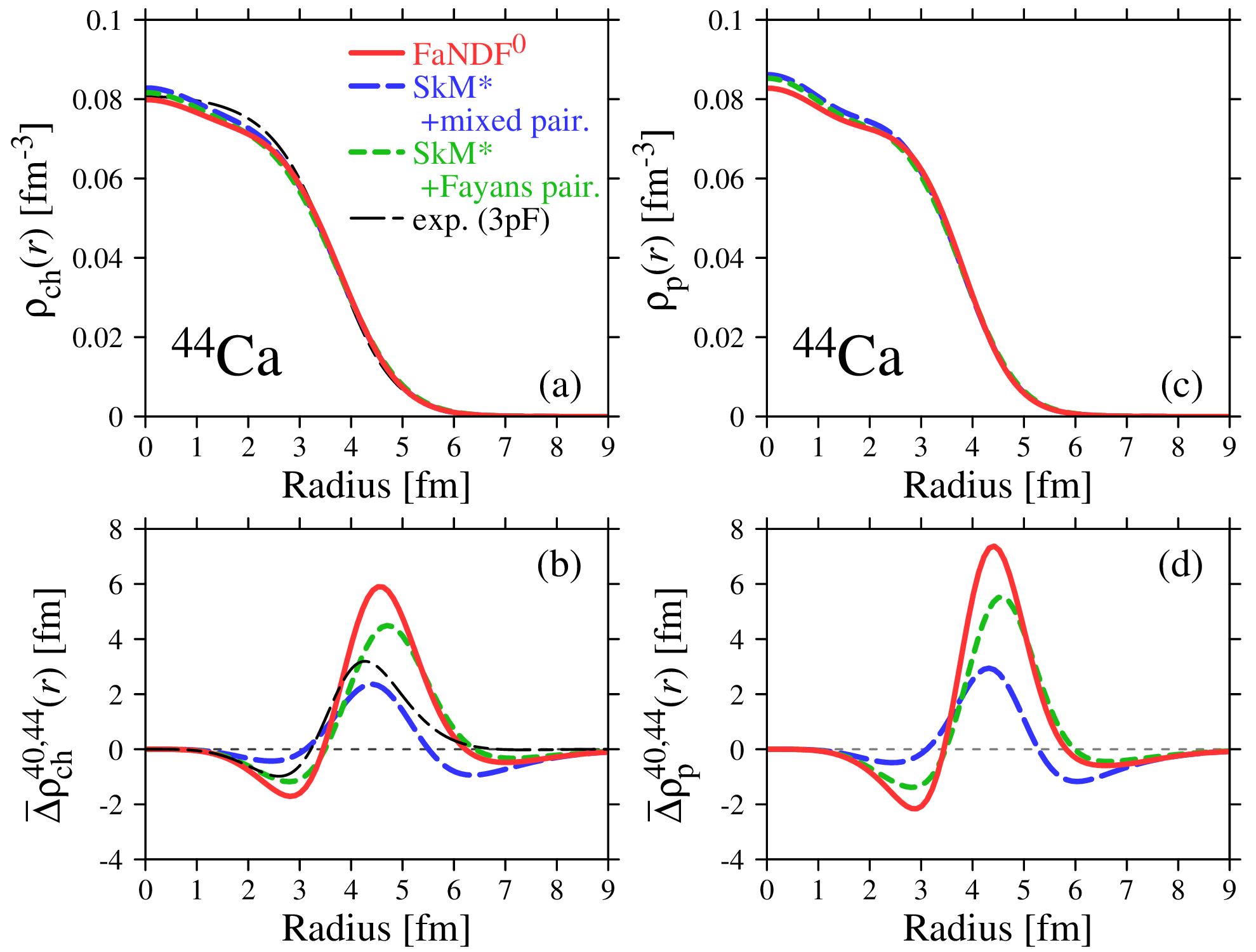}
\caption{Charge density distributions and point-proton density distributions of $^{44}$Ca and those differences from $^{40}$Ca, 
$\bar{\Delta}\rho_{{\rm ch/p}}^{40,44}(r)$, calculated with FaNDF$^0$ and SkM$^\ast$, 
and experimental data in the 3pF model.}
\label{charge.density.44Ca}
\end{center}
\end{figure}

Figures~\ref{charge.density.44Ca}(a)(b) show the charge density distribution of $^{44}$Ca, $\rho_\mathrm{ch}(r,\,^{44}\mathrm{Ca})$, and its difference from that of $^{40}$Ca,
$\bar{\Delta}\rho_{\rm ch}^{A_0,A}(r)$ defined in Eq.~\eqref{eq:diff_rch} with $A_0=40$.
The integrated value of $\bar{\Delta}\rho_{\rm ch}^{40,A}$ over $r$ is equivalent to $\delta\langle r^2\rangle_{\rm ch}^{40,A}$ plotted in Fig.~\ref{isotope.shift}; thus
$\bar{\Delta}\rho_{\rm ch}^{40,A}$ shows the contribution to $\delta\langle r^2\rangle_{\rm ch}^{40,A}$ as a function of $r$.
Experimental data on $\bar{\Delta}\rho_{\rm ch}^{40,A}$ is taken from Ref.~\cite{DEVRIES1987495}, which is extracted from the densities in terms of the three-parameter Fermi function (3pF) fitted to the charge form factors.
In $^{44}$Ca, neutrons gain the pair correlations.
In addition to the mixed-type pairing functional (\ref{mixed.pairing}),
the same pairing functional (\ref{FayansEDF.pairing}) as in the FaNDF$^0$~\cite{Reinhard2017} is applied to $^{44}$Ca in combination with the particle-hole part of SkM$^\ast$,
in order to investigate the influence of the pairing functional.
Although the results of $\rho_\mathrm{ch}(r,\mbox{$^{44}$Ca})$ are not well distinguished [Fig.~\ref{charge.density.44Ca}(a)],
those of $\bar{\Delta}\rho_{\rm ch}^{40,44}$, the differences of the charge densities from $^{40}$Ca, are resolved [Fig.~\ref{charge.density.44Ca}(b)].
For $\bar{\Delta}\rho_{\rm ch}^{40,44}$,
all the calculations give dips in the inner region of nuclei and peaks in the surface region, which are qualitatively consistent with the experimental data (black dash-dot-dotted line).
It seems that the SkM$^\ast$\,+\,mixed pairing (blue long-dashed line) has the peak height closer to the experimental data, but has a large negative contribution of $\bar{\Delta}\rho_{\rm ch}^{40,44}$ in the outer region around $r\sim 5$--8 fm, which offsets the positive contribution to $\langle r^2\rangle_{\rm ch}$ in the surface region.
The FaNDF$^0$ (red solid line) has a higher peak than the experimental data, and has a negative contribution in the outer region,
which the experimental data does not have.
As a consequence, FaNDF$^0$ well reproduces the enhancement of the charge radius at $^{44}$Ca.
It is noted that the experimental data on $\bar{\Delta}\rho_{\rm ch}^{40,44}$ could contain systematic errors originating from the 3pF model.
When the pairing functional is changed from the mixed-type one to the Fayans one (green dashed line) combined with the particle-hole part of SkM$^\ast$, the charge density difference becomes similar to that of FaNDF$^0$. 
This indicates that the Fayans pairing (\ref{FayansEDF.pairing}) is responsible for the parabolic behavior of the charge radius in $^{40\textrm{--}48}$Ca.

\begin{figure}[tb]
\begin{center}
\includegraphics[width=0.450\textwidth,keepaspectratio]
{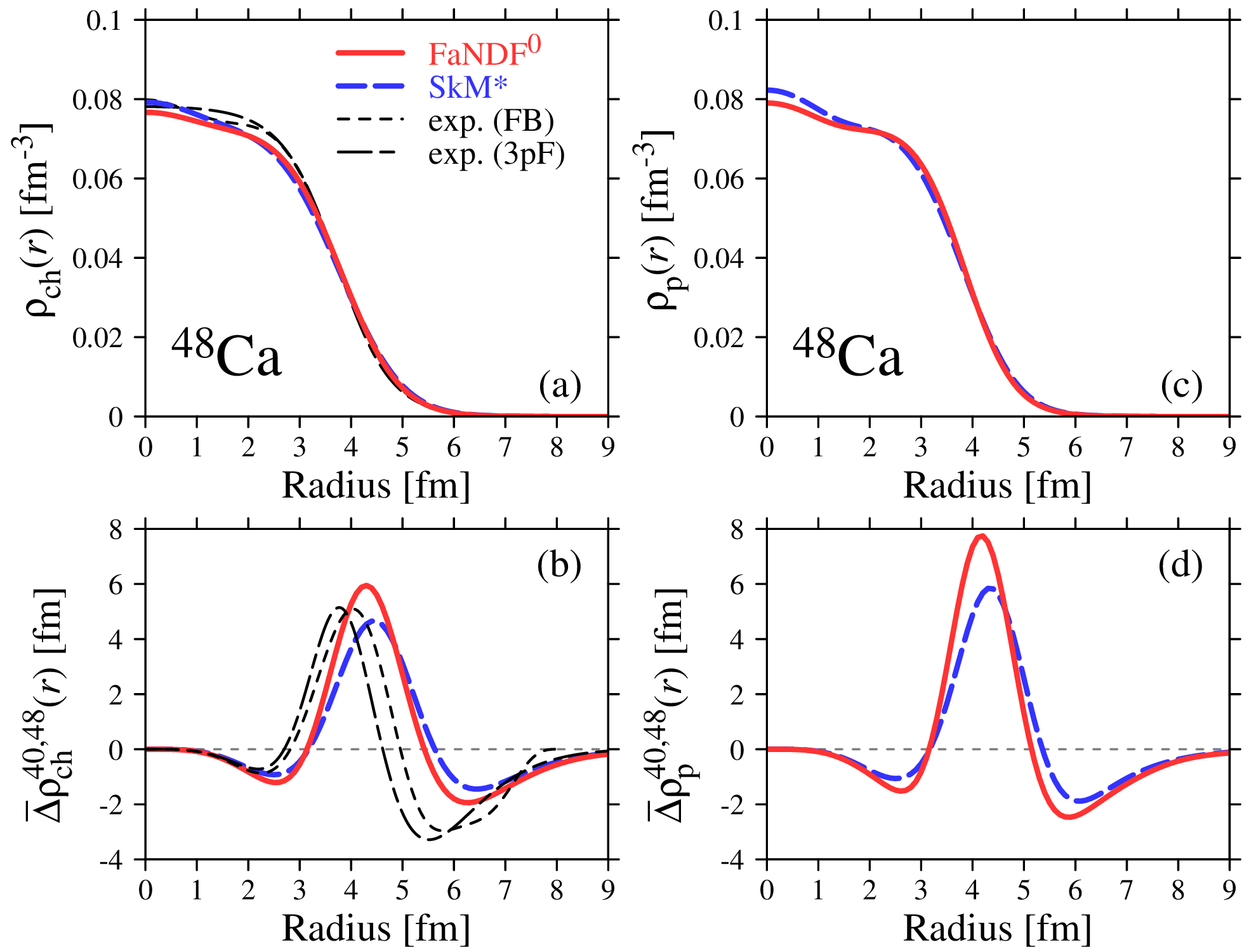}
\caption{Charge and point-proton density distributions of $^{48}$Ca $\rho_{{\rm ch/p}}(r,\,^{48}\mathrm{Ca})$ and those differences from $^{40}$Ca, $\bar{\Delta}\rho_{{\rm ch/p}}^{40,48}(r)$, calculated with FaNDF$^0$ and SkM$^\ast$, 
and experimental data for charge radius distributions, analyzed in the Fourier-Bessel (FB) expansion and the 3pF model.}
\label{charge.density.48Ca}
\end{center}
\end{figure}

Figures~\ref{charge.density.48Ca}(a)(b) show the calculated charge density distributions of $^{48}$Ca, $\rho_\mathrm{ch}(r,\,^{48}\mathrm{Ca})$, and their differences from those of $^{40}$Ca, $\bar{\Delta}\rho_{\rm ch}^{40,48}$.
They are also compared with the experimental data~\cite{DEVRIES1987495} which are analyzed by the Fourier-Bessel (FB) expansion and by the 3pF model.
The small deviation between the data extracted from the FB and the 3pF models suggests that the model dependence in the experimental analyses is not quite significant.
No pairing correlation is activated in the ground state of $^{48}$Ca in the present calculations.
As at $^{44}$Ca, both FaNDF$^0$ and SkM$^\ast$ produce similar charge density distributions $\rho_\mathrm{ch}(r)$ and are consistent with the experimental data [Fig.~\ref{charge.density.48Ca}(a)]. 
Even for $\bar{\Delta}\rho_{\rm ch}^{40,48}$ in Fig.~\ref{charge.density.48Ca}(b), the differences of the charge density distributions, the results of FaNDF$^0$ and SkM$^\ast$ resemble each other, having dips at $r=2\textrm{--}3$~fm, peaks at the surface region $r=4\textrm{--}5$~fm, and sizable negative contributions in the outer region.
These are qualitatively consistent with experimental data, although the positions of the dips and the peaks are located outward by $0.5\textrm{--}1$~fm compared to the experimental data.
The negative $\bar{\Delta}\rho_{\rm ch}^{40,48}$ values in the outer regions result in the almost equal charge radii between $^{40}$Ca and $^{48}$Ca viewed in Fig.~\ref{isotope.shift}.
On the other hand, Fig.~\ref{charge.density.44Ca}(b) shows that the dip of $\bar{\Delta}\rho_{\rm ch}^{40,44}$ in the outer region is quite shallow with FaNDF$^0$, which is consistent with the experimental data and produces the parabola structure of charge radius, while SkM$^\ast$ provides sizably negative values as in $\bar{\Delta}\rho_{\rm ch}^{40,48}$.

Although the corrections presented in Appendix~\ref{sec:chargedensity} have been taken into account,
$\rho_\mathrm{ch}(r)$ is dominated by the point-proton density $\rho_\mathrm{p}(r)$.
Figures~\ref{charge.density.44Ca}(c)(d) and \ref{charge.density.48Ca}(c)(d) display $\rho_\mathrm{p}(r)$ of $^{44,48}$Ca and their differences from $^{40}$Ca
defined in Eq.~\eqref{eq:diff_rhop}.
The above discussions for the charge density distributions hold for the proton density distributions, 
although the amplitudes of $\bar{\Delta}\rho_{\rm ch}^{40,A}(r)$ tend to be smaller than those of $\bar{\Delta}\rho_{\rm p}^{40,A}(r)$.
The same holds for the $N<20$ and $N>28$ regions.
In the following, we investigate the proton density distributions $\rho_\mathrm{p}(r)$ and the point-proton radii $\langle r^2\rangle_{\rm p}$, rather than $\rho_\mathrm{ch}(r)$ and $\langle r^2\rangle_{\rm ch}$, because they are more transparent.

\begin{figure}[tb]
\begin{center}
\includegraphics[width=0.480\textwidth,keepaspectratio]{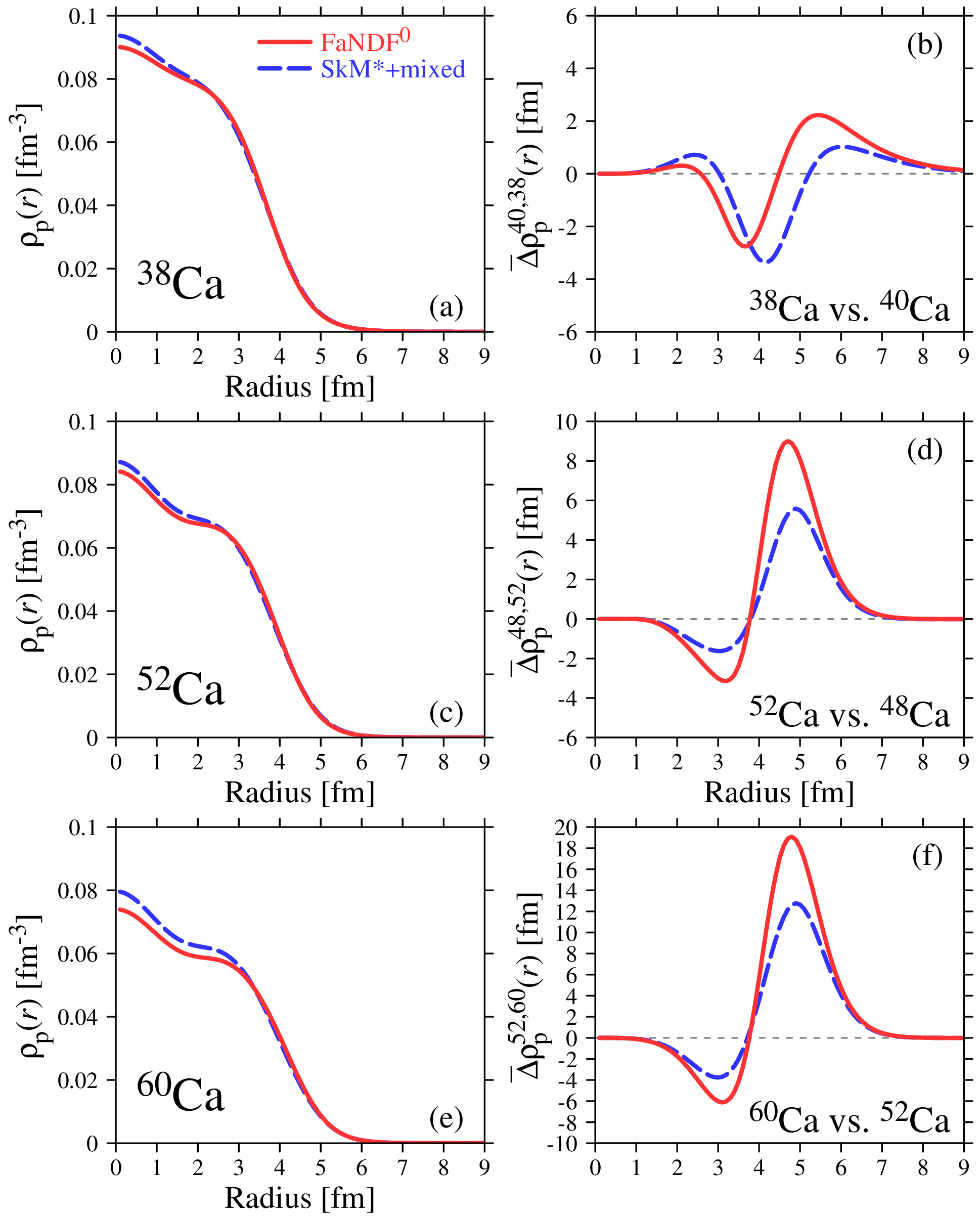}
\caption{Proton density distributions of (a) $^{38}$Ca, (b) $^{52}$Ca, and (c) $^{60}$Ca calculated with FaNDF$^0$ and SkM$^\ast$,
  and their difference from the reference nuclei, which is taken to be $^{40}$Ca for $^{38}$Ca in (b), $^{48}$Ca for $^{52}$Ca ($^{60}$Ca) in (d) [(f)].}
\label{proton.density.38Ca.52Ca.60Ca}
\end{center}
\end{figure}

We turn to the neutron-deficient nucleus $^{38}$Ca. 
As has been shown in Fig.~\ref{isotope.shift}, the measured charge radius of $^{38}$Ca is smaller than that of $^{40}$Ca, $\delta \langle r^2 \rangle^{38,40}_\mathrm{ch}= -0.080$~fm$^2$. 
SkM$^\ast$ and M3Y-P6 reproduce $\delta \langle r^2 \rangle^{38,40}_\mathrm{ch}$ well, while FaNDF$^0$ overestimates it, $\delta \langle r^2 \rangle^{38,40}_\mathrm{ch}= +0.095$~fm$^2$.
We analyze the differential radii via the density distributions.
Figures~\ref{proton.density.38Ca.52Ca.60Ca}(a)(b) show the proton density distributions and those differences from $^{40}$Ca of Eq.~\eqref{eq:diff_rhop}, calculated with FaNDF$^0$ and SkM$^\ast$.
Both FaNDF$^0$ and SkM$^\ast$ provide $\bar{\Delta}\rho_{\rm p}^{40,38}$ with dips around $r=4\,\mathrm{fm}$.
However, FaNDF$^0$ has a significant peak in the outer region, washing out the negative contribution of the dip to $\delta \langle r^2 \rangle^{38,40}_\mathrm{ch}$, whereas the corresponding peak stays short in the SkM$^\ast$ result.
Thus, similar to the $^{44}$Ca case, the $N$-dependence of the density in the outer region makes the difference in the charge radii between FaNDF$^0$ and SkM$^\ast$.

For the neutron-rich nuclei with $N>28$, the charge radius increases monotonically both experimentally and theoretically, as shown in Fig.~\ref{isotope.shift}.
The difference in the neutron pairing functionals does not make a qualitative difference for the charge radii.
The distributions of the proton density swelling shown in Figs.~\ref{proton.density.38Ca.52Ca.60Ca}(d)(f) are similar between FaNDF$^0$ and SkM$^\ast$ despite the difference in their amplitudes. 

\subsection{Orbital decomposition}\label{subsec:orbital}

\begin{figure}[tb]
\begin{center}
\includegraphics[width=0.400\textwidth,keepaspectratio]{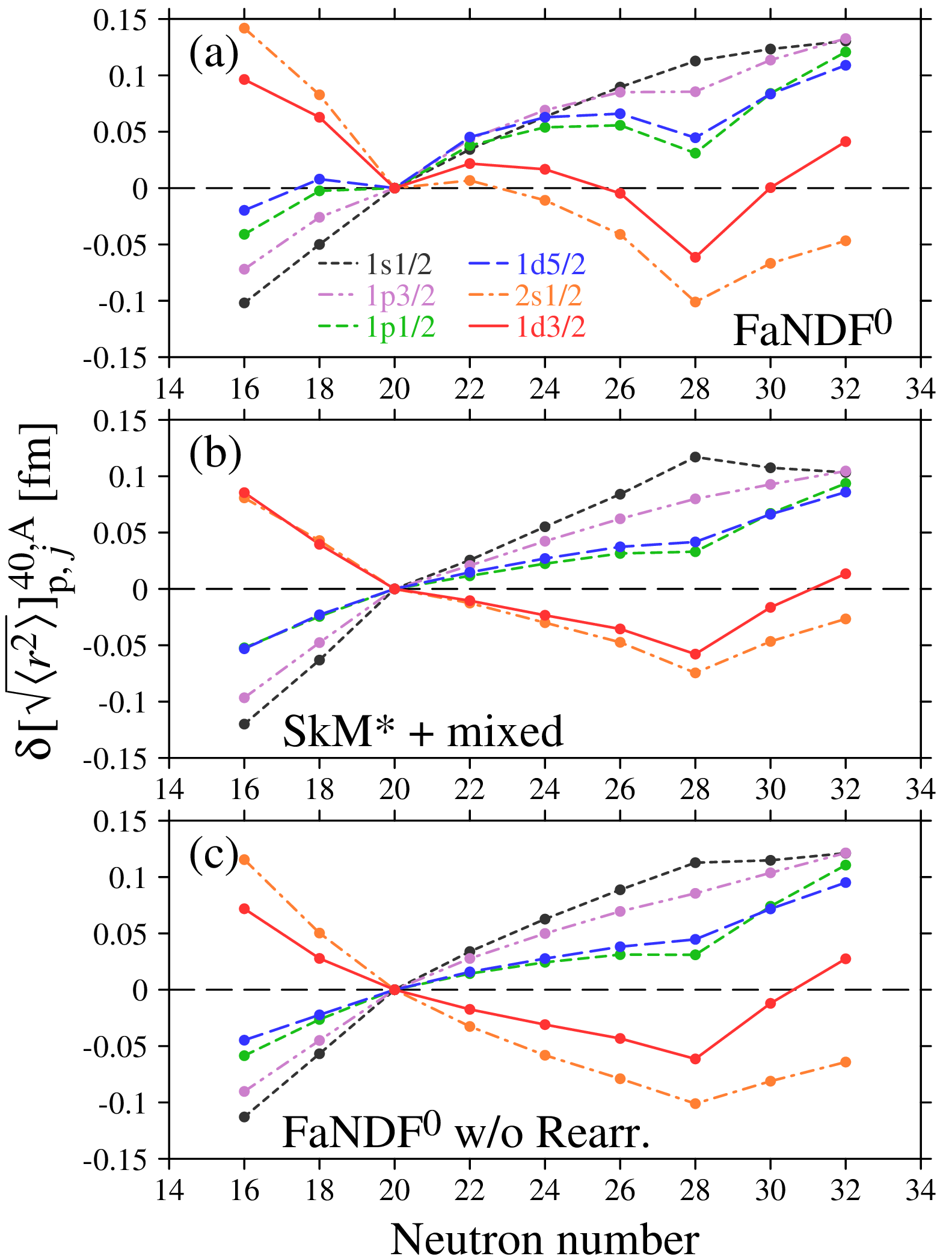}
\caption{Differences of proton singe-particle radii calculated with (a) FaNDF$^0$, (b) SkM$^\ast$, and (c) FaNDF$^0$ with omitting the rearrangement term. See the text for details.}
\label{diff.proton.spRrms}
\end{center}
\end{figure}

Because $Z=20$ is a magic number, the proton pair correlation is not active in all the calculations presented here.
The proton radius and density distribution can be decomposed into the contributions of the proton s.p. orbits that are filled at the $Z=20$ magic number.
The orbital decomposition may provide additional information on the $N$-dependence of the radius,
telling us the influence of the neutron occupation on specific orbits.
To identify the orbital contribution to the differential proton radius,
we show the differences of the radii of proton s.p. orbit from those of $^{40}$Ca, $\delta\big[\sqrt{\langle r^2\rangle}\big]^{40,A}_{\mathrm{p},j}$ of Eq.~\eqref{eq:diff_rpj} as functions of $N$, calculated with FaNDF$^0$ in Fig.~\ref{diff.proton.spRrms}(a).
The corresponding quantities with SkM$^\ast$ are shown in Fig.~\ref{diff.proton.spRrms}(b).
Significant difference is found between them, corresponding to the difference in $\delta\langle r^2\rangle^{40,A}_\mathrm{ch}$.
At $N=22$, all proton orbits have larger radii than those of $^{40}$Ca in the FaNDF$^0$ result.
At $^{44}$Ca, $1d_{3/2}$ and $2s_{1/2}$ orbits shrink slightly while the other proton orbits keep expanding. 
The differential radii of proton orbit are convex upward in $^{40\textrm{--}48}$Ca irrespective of $j$.
All the orbits except $2s_{1/2}$ have positive contributions to $\delta \langle r^2\rangle^{40,44}_\mathrm{p}$,
resulting in the proton and charge radii at $^{44}$Ca larger than $^{40}$Ca.
Therefore, many orbits contribute coherently to the swelling of proton and charge radii at $^{44}$Ca.
In particular, the $sd$-shell orbits, \textit{i.e.}, $1d_{5/2}$, $1d_{3/2}$ and $2s_{1/2}$, predominantly contribute to the convexity.
Note that the values in Fig.~\ref{diff.proton.spRrms} should be multiplied by the occupation number, which is equal to the degeneracy of the orbit $(2j+1)$, to obtain the differential radii $\delta\langle r^2\rangle^{A_0,A}_\mathrm{ch/p}$,
enhancing the contribution of $1d_{5/2}$.
At $^{48}$Ca, the proton $1d_{3/2}$ and $2s_{1/2}$ orbits shrink from those of $^{40}$Ca by 0.061\,fm and 0.101\,fm, canceling the expansion of the other orbits and leading to $\delta\langle r^2\rangle^{40,48}_\mathrm{ch}\approx 0$.
Meanwhile, Fig.~\ref{diff.proton.spRrms}(b) shows that SkM$^\ast$ gives almost linear $\delta\big[\sqrt{\langle r^2\rangle}\big]^{40,A}_{\mathrm{p},j}$ in $^{40\textrm{--}48}$Ca.
The proton radii calculated with SkM$^\ast$ do not vary significantly in $^{40\textrm{--}48}$Ca because the shrinking $1d_{3/2}$ and $2s_{1/2}$ orbits cancel the spread of the other orbits,
as observed for the charge radii in Fig.~\ref{isotope.shift}.
When we employ FaNDF$^0$ but with omitting the rearrangement terms of the neutron pairing correlation in the central potential, $\delta \mathcal{E}_\mathrm{pair} / \delta \rho_t$, the behavior of the proton-radius difference becomes similar to 
those of SkM$^\ast$\,+\,mixed pairing, as shown in Fig.~\ref{diff.proton.spRrms}(c) and Ref.~\cite{Reinhard2017}.
The rearrangement term 
is crucial for reproducing the parabolic structure of the proton and charge radii in $^{40\textrm{--}48}$Ca.
Note that the $1s_{1/2}$ and $2s_{1/2}$ orbits may easily mix each other in the mean-field calculations under the spherical symmetry via a small off-diagonal component.
The summed contribution from the $1s_{1/2}$ and $2s_{1/2}$ orbits stays small for all cases in Fig.~\ref{diff.proton.spRrms}.

Turning to $^{38}$Ca, we have found that $\delta\big[\sqrt{\langle r^2\rangle}\big]_{\mathrm{p},1d_{5/2}}^{40,38}$ is positive with FaNDF$^0$, in contrast to that with SkM$^\ast$ and without the rearrangement term of the pairing.
Also, $\delta\big[\sqrt{\langle r^2\rangle}\big]_{\mathrm{p},1p_{1/2}}^{40,38}$ is vanishing with FaNDF$^0$,
significantly contributing to the positive $\delta\langle r^2\rangle_\textrm{ch}^{40,38}$ in Fig.~\ref{isotope.shift},
while sizably negative with SkM$^\ast$ and without the rearrangement term.

\begin{figure}[tb]
\begin{center}
\includegraphics[width=0.480\textwidth,keepaspectratio]{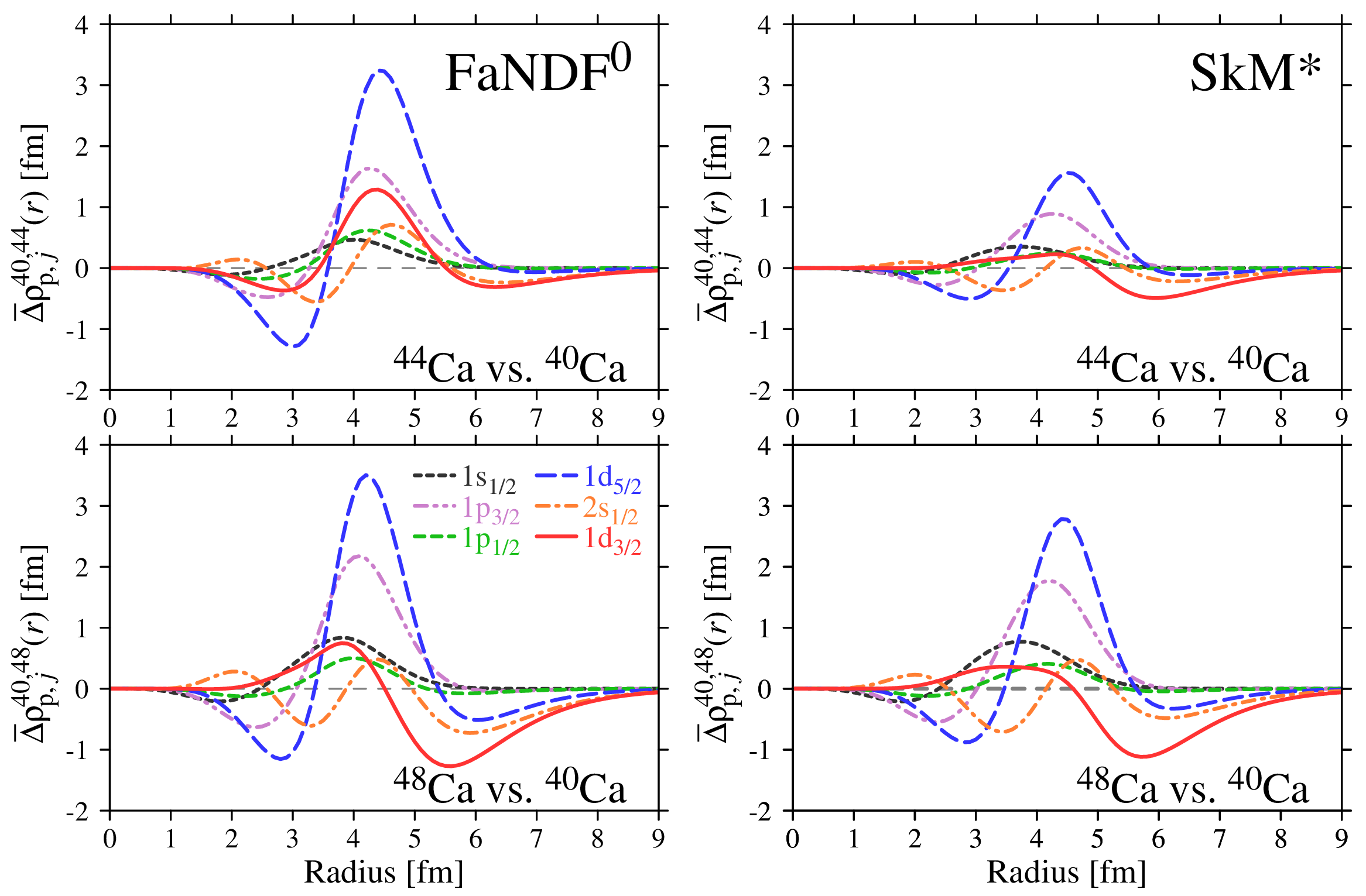}
\caption{Differences of proton density distributions of $^{44,48}$Ca from $^{40}$Ca, decomposed to proton orbitals, 
$\bar{\Delta}\rho_{{\rm p},j}^{40,44}(r)$
 and $\bar{\Delta}\rho_{{\rm p},j}^{40,48}(r)$. FaNDF$^0$ and SkM$^\ast$ are used.}
\label{proton.density.decomp.44Ca.48Ca}
\end{center}
\end{figure}

We then analyze the radial contribution of each proton orbit to the differential proton radii.
Figure~\ref{proton.density.decomp.44Ca.48Ca} shows the differences of proton density distributions of $^{44,48}$Ca from $^{40}$Ca decomposed into the proton s.p. orbits,
$\bar{\Delta}\rho_{{\rm p},j}^{40,44}(r)$ and $\bar{\Delta}\rho_{{\rm p},j}^{40,48}(r)$ of Eq.~\eqref{eq:diff_rhopj}, calculated with the FaNDF$^0$ and SkM$^\ast$.
At $^{44}$Ca with FaNDF$^0$, all proton density differences except the $s_{1/2}$ orbits have shallow dips in the inner region and high peaks in the surface region. 
The $1d_{3/2}$ and $2s_{1/2}$ orbits have small negative contributions at the outer region.
These radial structure of $\bar{\Delta}\rho_{{\rm p},j}^{40,44}(r)$ accounts for $\delta\big[\sqrt{\langle r^2\rangle}\big]^{40,44}_{\mathrm{p},j}$.
We see some differences between $\bar{\Delta}\rho_{{\rm p},j}^{40,48}(r)$ and $\bar{\Delta}\rho_{{\rm p},j}^{40,44}(r)$.
Compared with $\bar{\Delta}\rho_{{\rm p},j}^{40,44}(r)$,
the peak positions of $\bar{\Delta}\rho_{{\rm p},j}^{40,48}(r)$ of each orbit at the surface region shift inward by 0.2--0.3\,fm, and the peak heights are also changed by 0.1--0.5\,fm. 
The most prominent difference is that $1d_{3/2}$ and $2s_{1/2}$ have sizable negative contributions in the outer region at $^{48}$Ca,
which shrinks the proton s.p. radius of $1d_{3/2}$ and $2s_{1/2}$ orbits by 0.061 fm and 0.101 fm, as shown in Fig.~\ref{diff.proton.spRrms}(a).
The $1d_{5/2}$ orbit also has a non-negligible negative contribution in the outer region. 
These shrinks of the proton orbits make the proton and charge radii of $^{48}$Ca smaller than those of $^{44}$Ca.

With SkM$^\ast$, $\bar{\Delta}\rho_{{\rm p},j}^{40,48}(r)$ have similar shapes to those with FaNDF$^0$.
At $^{44}$Ca, $\bar{\Delta}\rho_{{\rm p},j}^{40,44}(r)$ have small amplitudes compared to those with FaNDF$^0$.
The $1d_{3/2}$ and $2s_{1/2}$ orbits give negative density differences,
which offset the positive contributions of the $1s_{1/2}$, $1p_{3/2}$, $1p_{1/2}$, and $1d_{5/2}$ orbits, resulting in almost the equal charge radii of $^{40,44}$Ca. 
The shapes of $\bar{\Delta}\rho_{{\rm p},j}^{40,44}(r)$ are analogous to those of $\bar{\Delta}\rho_{{\rm p},j}^{40,48}(r)$ but with different scales.
Therefore, the charge radii of $^{40,44,48}$Ca calculated with SkM$^\ast$ are close.

\begin{figure}[tb]
\begin{center}
\includegraphics[width=0.480\textwidth,keepaspectratio]{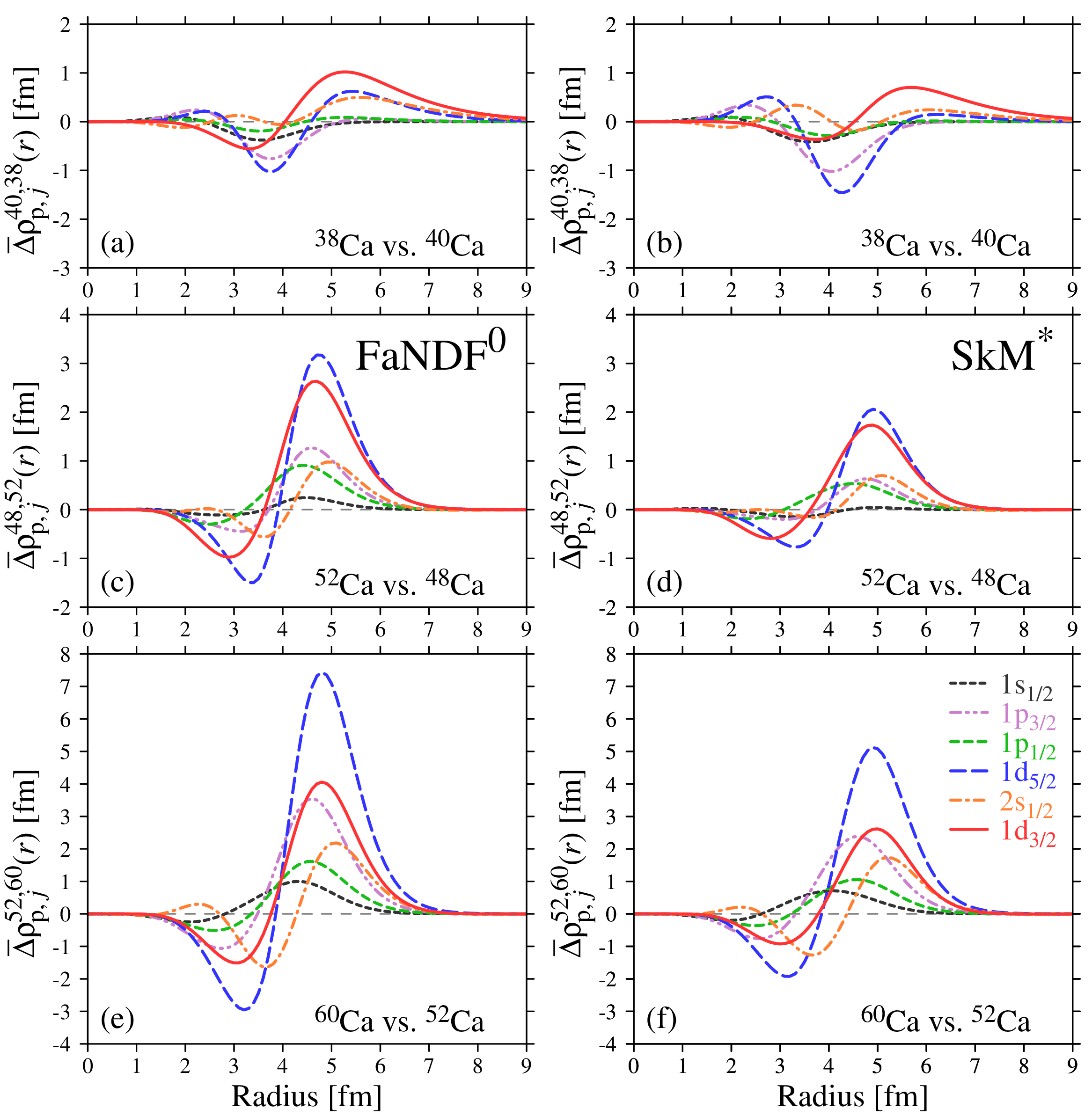}
\caption{Differences of proton density distributions between (a)(d) $^{40}$Ca and $^{38}$Ca, (b)(e) $^{48}$Ca and $^{52}$Ca, and (c)(f) $^{52}$Ca and $^{60}$Ca, decomposed to proton orbitals. FaNDF$^0$ and SkM$^\ast$ are used. }
\label{proton.density.decomp.38Ca.52Ca.60Ca}
\end{center}
\end{figure}

Figures~\ref{proton.density.decomp.38Ca.52Ca.60Ca}(a)(b) show $\bar{\Delta}\rho_{{\rm p}}^{40,38}(r)$ decomposed to proton orbits.
The large positive $\bar{\Delta}\rho_{{\rm p}}^{40,38}(r)$ in the outer region with FaNDF$^0$ viewed in Fig.~\ref{proton.density.38Ca.52Ca.60Ca}(b) comes mainly from the $1d_{5/2}$, $1d_{3/2}$ and $2s_{1/2}$ orbits.
It has a similarity with $\bar{\Delta}\rho_{{\rm p}}^{40,44}(r)$ in the significant role of the $sd$-shell orbits.

From the decomposition of proton-density swellings in Figs.~\ref{proton.density.decomp.38Ca.52Ca.60Ca}(c)\,--\,(f), it is seen that all proton orbits have coherent positive contributions at the surface and outer regions in the neutron-rich Ca nuclei.

\subsection{Difference in potentials}

\begin{figure}[tb]
\begin{center}
\includegraphics[width=0.480\textwidth,keepaspectratio]{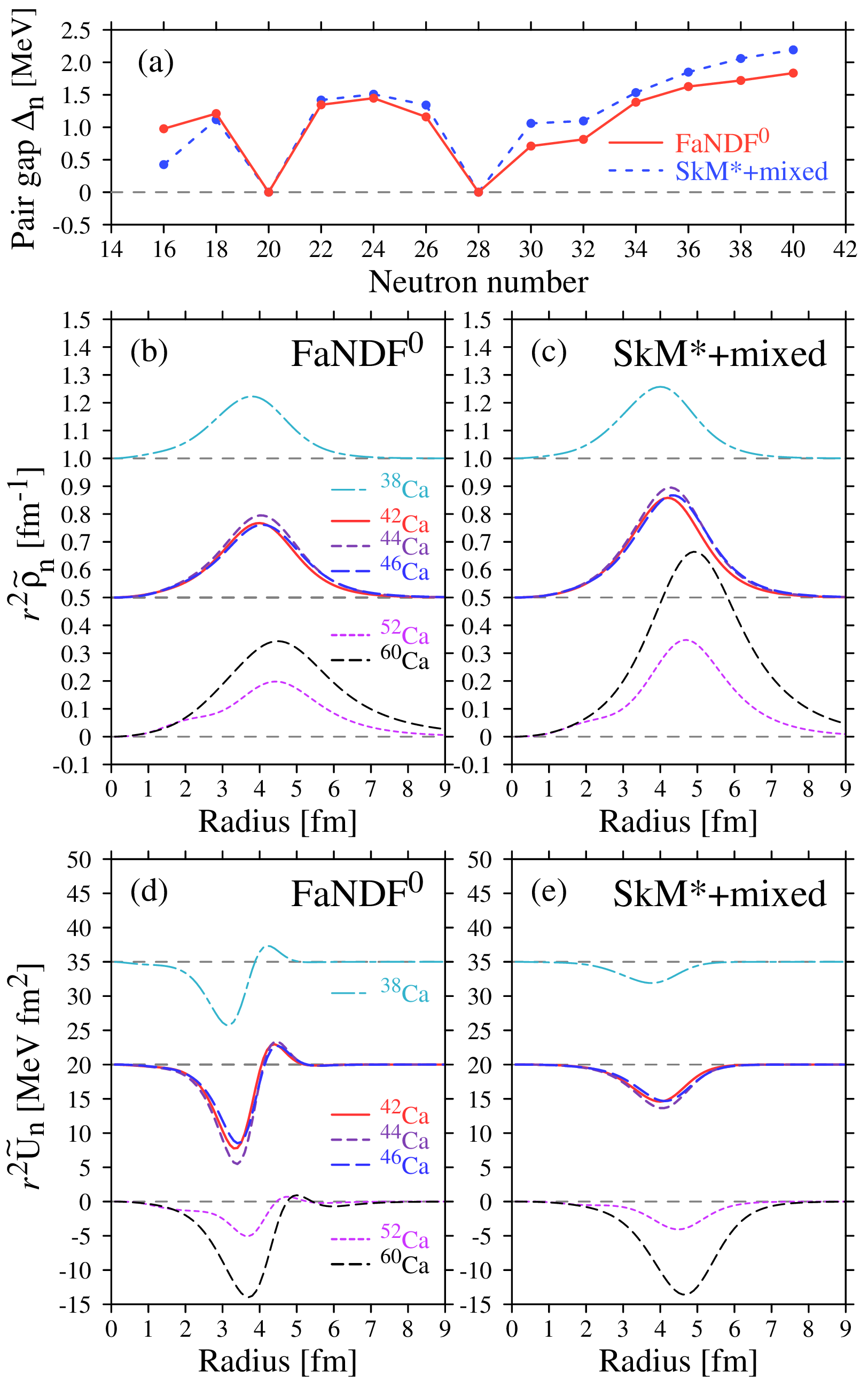}
\caption{Neutron pair gaps $\Delta_{\mathrm{n}}$, neutron pair densities $r^2 \tilde{\rho}_{\mathrm{n}}$, and neutron pairing potentials $r^2 \tilde{U}_{\mathrm{n}}$ of even Ca isotopes, calculated with FaNDF$^0$ and SkM$^\ast$.
In (b)\,--\,(e), the curves are grouped and vertically shifted to accommodate them.
}
\label{Upairn}
\end{center}
\end{figure}

In the Fayans EDF results, the neutron pairing correlation is the key to producing the parabola structure of charge radius in $^{40\textrm{--}48}$Ca. 
Figure~\ref{Upairn} compares the calculated neutron pair gaps $\Delta_{\mathrm{n}}$, neutron pair density $r^2\tilde{\rho}_{\mathrm{n}}$, and neutron pair potentials $r^2 \tilde{U}_{\mathrm{n}}$ calculated with FaNDF$^0$ and SkM$^\ast$\,(+mixed pairing) in the Ca isotopes. 
The FaNDF$^0$ and SkM$^\ast$ functionals produce similar pair gaps, and the neutron pair density distributions $r^2 \tilde{\rho}_{\mathrm{n}}$ are not so different as well, 
while the peaks of $r^2 \tilde{\rho}_{\mathrm{n}}$ of SkM$^\ast$ are higher and located outward by 0.2--0.3 fm than those of FaNDF$^0$ in $^{42\textrm{--}46}$Ca.
However, there is a notable difference in the neutron pair potentials $r^2\tilde{U}_{\mathrm{n}}$.
This difference should come from the pairing functionals in Eqs.~(\ref{mixed.pairing}) and (\ref{FayansEDF.pairing}).
While SkM$^\ast$ gives valleys of $r^2\tilde{U}_{\textrm{n}}$ at $r \sim 4.1$\,fm,
FaNDF$^0$ yields valleys at $r \sim 3.4$\,fm and bumps at $r\sim 4.5$\,fm in $^{42\textrm{--}46}$Ca, with nodes at the surface region $r\sim 4$\,fm
that are not produced by SkM$^\ast$.
With FaNDF$^0$, $r^2\tilde{U}_{\mathrm{n}}$ almost vanishes soon beyond the bump, more rapidly than with SkM$^\ast$.
It is remarked that this difference is also found in the $N<20$ region.
In the $N>28$ region, a similar difference exists but is less conspicuous.

\begin{figure}[tb]
\begin{center}
\includegraphics[width=0.480\textwidth,keepaspectratio]{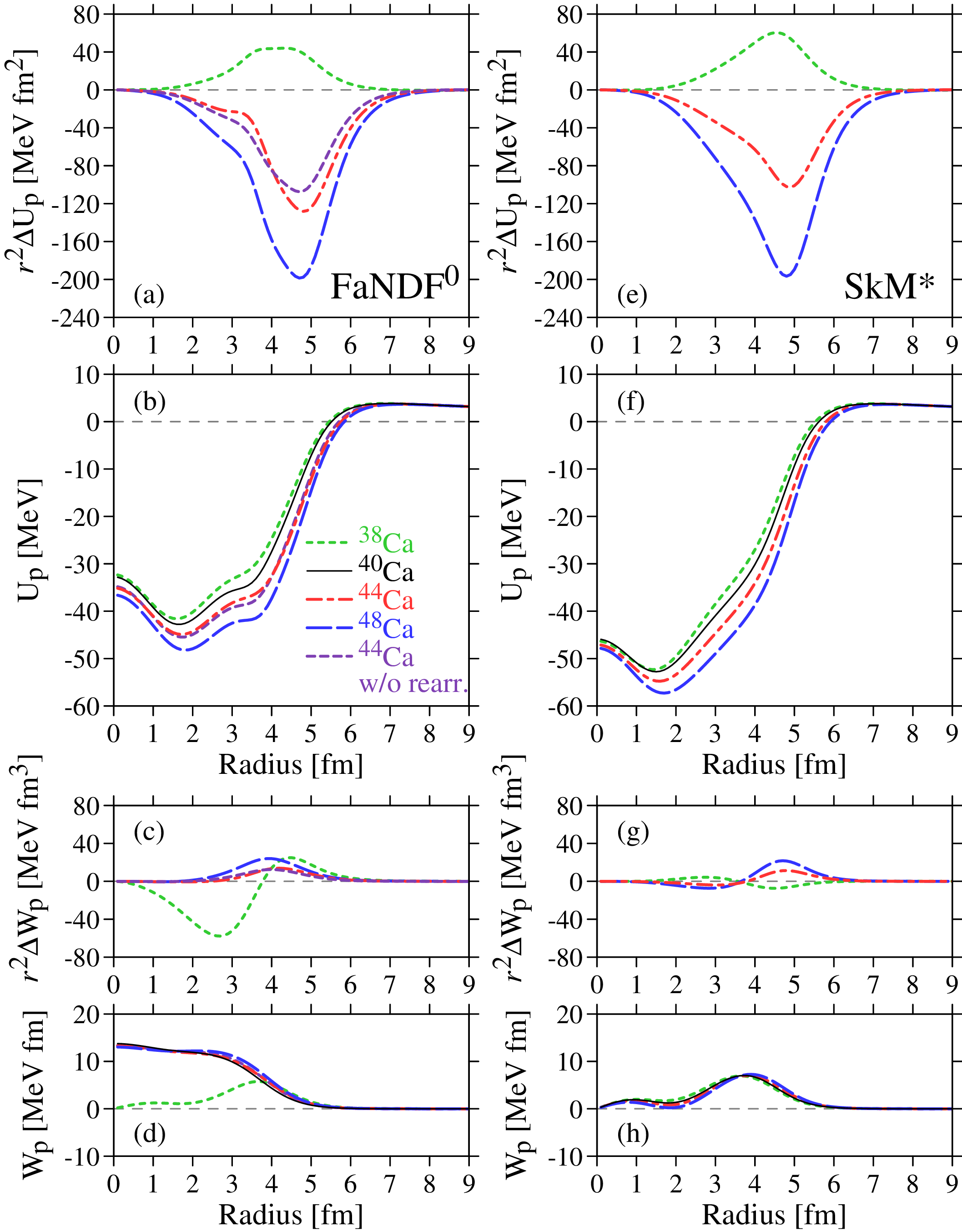}
\caption{Proton central potential $U_{\mathrm{p}}$ and LS potential $W_{\mathrm{p}}$ calculated with FaNDF$^0$ and SkM$^\ast$, and 
those differences from $^{40}$Ca.}
\label{Up}
\end{center}
\end{figure}

The neutron pairing correlations affect the proton potential through the rearrangement term $\delta \mathcal{E}_\mathrm{pair}/\delta \rho_{\mathrm{p}}$.
If we omit the rearrangement term  $\delta \mathcal{E}_\mathrm{pair}/\delta \rho_{\mathrm{p}}$ in the proton central potential, the differences of the proton s.p. radius calculated with FaNDF$^0$ become similar to those calculated with ${\rm SkM}^\ast + {\rm mixed}$ pairing,
as shown in Fig.~\ref{diff.proton.spRrms}(c).
Figure~\ref{Up} shows the proton central potential $U_{\mathrm{p}}$ and the proton spin-orbit (LS) potential 
$W_{\mathrm{p}}$ of $^{40,44,48}$Ca, the differences of those potentials from $^{40}$Ca, 
$r^2 \Delta U_{\mathrm{p}}^{40,A} := r^2 \left[ U_\mathrm{p} (^A\mathrm{Ca}) - U_\mathrm{p} (^{40}\mathrm{Ca})\right]$
and $r^2 \Delta W_{\mathrm{p}}^{40,A} := r^2 \left[ W_\mathrm{p} (^A\mathrm{Ca}) - W_\mathrm{p} (^{40}\mathrm{Ca})\right]$.
The proton central potentials with omitting the rearrangement term at $^{44}$Ca are also plotted for reference.
Rather than $U_{\mathrm{p}}$ itself, the differences between FaNDF$^0$ and SkM$^\ast$ in $r^2 \Delta U_{\mathrm{p}}^{40,A}$ are more relevant to $\delta\langle r^2\rangle^{40,A}_\mathrm{ch/p}$ under discussion.
With FaNDF$^0$, $r^2\Delta U_{\mathrm{p}}^{40,44}$ [red dot-dashed line in Fig.~\ref{Up}(a)] suddenly changes its slope at $r \sim 3.5$\,fm and has a valley at $r\sim 4.7$\,fm, corresponding to the bottom and the top of $r^2\tilde{U}_{\mathrm{n}}$ in Fig.~\ref{Upairn}.
This structure of $r^2\Delta U_{\mathrm{p}}^{40,44}$ makes the dip and the peak of $\bar{\Delta}\rho_{{\rm ch/p}}^{40,44}(r)$ more conspicuous than those with SkM$^\ast$ in Fig.~\ref{charge.density.44Ca}.
Beyond the valley region, $r^2\Delta U_{\mathrm{p}}^{40,44}$ damps more slowly with FaNDF$^0$ than with SkM$^\ast$,
by which the dip of $\bar{\Delta}\rho_{\rm p}^{40,44}(r)$ in the outer region becomes shallow, as observed in Fig.~\ref{charge.density.44Ca}.
If the rearrangement term $\delta \mathcal{E}_\mathrm{pair} / \delta \rho_{\mathrm{p}}$ is omitted, the slope change in $r^2 \Delta U_{\mathrm{p}}^{40,44}$ becomes milder (purple dashed line).
The proton density difference resembles that of SkM$^\ast$, 
and, as is shown in Fig.~\ref{diff.proton.spRrms}, 
the proton s.p. radius differences also become similar to those of SkM$^\ast$.
For the LS potentials, $W_{\mathrm{p}}$ of $^{40,44,48}$Ca are similar to one another, and 
$\Delta W_{\mathrm{p}}^{40,A}$ has much smaller amplitudes than $\Delta U_{\mathrm{p}}^{40,A}$ has. 
Thus, the influence of the LS potential $\Delta W_{\mathrm{p}}^{40,A}$ on the proton radii and the charge radii is negligible.

The neutron pair density $r^2\tilde{\rho}_\mathrm{n}$ and the neutron pair potential $r^2\tilde{U}_\mathrm{n}$ in $^{38}$Ca are in parallel to those in $^{42\textrm{--}46}$Ca [Fig.~\ref{Upairn}(d)]. 
If we omit the rearrangement term  $\delta \mathcal{E}_\mathrm{pair}/\delta \rho_{\mathrm{p}}$ in the proton central potential with FaNDF$^0$, the calculated charge radius becomes close to the experimental data, $\delta \langle r^2 \rangle^{38,40}_\mathrm{ch}= -0.085$\,fm$^2$.
It further confirms that the neutron pairing functional of Fayans EDF, which looked favorable for the charge radii in $^{40\textrm{--}48}$Ca, simultaneously gives rise to the overestimation of the charge radius in $^{38}$Ca.

For $^{38}$Ca, $r^2 \Delta U_{\mathrm{p}}^{40,38}$ has the positive sign, opposite to $r^2 \Delta U_{\mathrm{p}}^{40,44}$.
Apart from the difference in the sign, $r^2 \Delta U_{\mathrm{p}}^{40,38}$ is connected to $r^2\tilde{U}_{\mathrm{n}}$ in Fig.~\ref{Upairn} analogously to $r^2\Delta U^{40,44}_\mathrm{p}$.
The positions of the sudden change in the slope and the peak of the $r^2\Delta U_\mathrm{p}^{40,38}$ correspond to the bottom and the top of $r^2\tilde{U}_\mathrm{n}$, also deriving $\bar{\Delta} \rho^{40,38}(r)$ in Fig.~\ref{proton.density.38Ca.52Ca.60Ca}.
With FaNDF$^0$, the slope of $r^2 \Delta U_{\mathrm{p}}^{40,38}$ changes at $r \sim 3.5$\,fm, starting a plateau, which is not observed in $r^2 \Delta U_{\mathrm{p}}^{40,38}$ with SkM$^\ast$.
The position of this slope change coincides with the minimum of $\bar{\Delta}\rho_{{\rm p}}^{40,38}(r)$.
Owing to the plateau structure, the peak of $r^2 \Delta U_{\mathrm{p}}^{40,38}$ is lower with FaNDF$^0$ than that with SkM$^\ast$,
yielding the shallower valley and the larger maximum at $r\sim 5$\,fm of $\bar{\Delta}\rho_{{\rm p}}^{40,38}(r)$.

\section{Summary and discussions}

We have investigated the charge and point-proton radii of the Ca nuclei in the DFT framework.
It has been argued that the Fayans EDF provides parabolic $N$-dependence in the differential charge radii in $20\leq N\leq 28$, which looks compatible with the experimental data,
due to its unique pairing channel.
By decomposing the charge and proton radii into the radial and orbital contributions and paying special attention to the differential values from a reference nucleus, mainly $^{40}$Ca, it is analyzed how the Fayans EDF provides specific $N$-dependence of the radii.
The results of the Fayans EDF (FaNDF$^0$) have been compared with those of SkM$^\ast$, which is a representative of EDFs having a usual pairing channel.

The characteristic $N$-dependence in the $20\leq N\leq 28$ region produced by the Fayans EDF has been traced back to the neutron pair potential, which has nodal behavior with a peak at smaller $r$ than the pair potential by the SkM$^\ast$, a dip and rapid damping.
It influences the central part of the proton single-particle potential as found via $r^2\Delta U_{\mathrm{p}}^{40,A}$,
and then the proton and charge densities.
In terms of the orbitals, influences on the $sd$-shell orbits, $1d_{3/2}$ and $2s_{1/2}$ in particular, are significant.
However, although the $N$-dependence of the charge radii has been compatible with the experimental data in $20\leq N\leq 28$ as a result,
the same mechanism gives rise to the erroneous enhancement of the charge radii in the $N<20$ region.
For the charge radii of $^{40\textrm{--}48}$Ca, there have also been studies suggesting roles of proton excitation across the $Z=20$ magic number.
For pinning down correct physics in the charge radii in the Ca isotopes, it is significant to discuss $^{36,38}$Ca simultaneously.
It would be interesting if radial or orbital information of the radii is obtained in future experiments.

We have investigated $N$-dependence of the charge and proton radii in the neutron-rich region as well.
Irrespective of EDFs with no qualitatively apparent distinction, a steady increase of the charge radii for increasing $N$ is predicted, although there is a quantitative difference.

\section*{Acknowledgments}
This work was supported by the JSPS KAKENHI
(Grant Nos. JP19KK0343 and JP20K03964).

\appendix 

\section{\texorpdfstring{FaNDF$^0$}{FaNDF0} parameter\label{sec:FaNDF0}}

Here the parameters of the FaNDF$^0$ \cite{Fayans1998} are summarized.
In this Appendix, we define all the parameters in terms of $\rho_{\rm sat}=0.16$ fm$^{-3}$
\footnote{We note that a different definition of the saturation density is used $\rho_{\rm sat} = 2\rho_0 = 0.16$ fm$^{-3}$ 
in the original reference \cite{Fayans1998}.}.
The Wigner-Seitz radius is given by
\begin{align}
 r_s = \left( \frac{3}{4\pi \rho_{\rm sat}}\right)^{\frac{1}{3}}  = 1.143 \,\,{\rm fm}.
\end{align}
The Fermi momentum and Fermi energies are
\begin{align}
k_F &= \left(\frac{3\pi^2 \rho_{\rm sat}}{2}\right)^{\frac{1}{3}} = 1.333\,\,{\rm fm}^{-1}, \\
 \epsilon_F &= \frac{\hbar^2 k_F^2}{2m}
= 36.846 \,\,{\rm MeV}.
\end{align}
In several references, the inverse density of state
is defined as 
\begin{align}
C_0 = \frac{4\epsilon_F}{3\rho_c} = 307.049\,\,{\rm MeV\,\,fm}^3.
\end{align}

The dimensionless parameters in FaNDF$^0$
are summarized in Table~\ref{table:FaNDF0}.

\begin{table}
\caption{Parameters for FaNDF$^0$. \label{table:FaNDF0}}
\begin{ruledtabular}
\begin{tabular}{cccc}
$a_+^{\rm v}$    &$-9.559$ & $\sigma$         & $1/3$  \\
$h^{\rm v}_{1+}$ & $0.633$ & $h_{\rm Coul} $  & $0.941$\\
$h^{\rm v}_{2+}$ & $0.131$ & $\kappa$         & $0.19$ \\
$a_-^{\rm v}$    & $4.428$ & $\kappa'$        & $0$    \\
$h^{\rm v}_{1-}$ & $0.250$ & $f^\xi_{\rm ex}$ &$-2.8$  \\
$h^{\rm v}_{2-}$ & $1.300$ & $h_+^\xi$        & $2.8$  \\
$a^{\rm s}_+$    & $0.600$ & $h^\xi_{\nabla}$ & $2.2$  \\
$h^s_{\nabla}$   & $0.440$ & $\gamma$         & $1$
\end{tabular}
\end{ruledtabular}
\end{table}

\section{Fayans potential\label{sec:potential}}

We here provide the explicit form for the central, spin-orbit, and pairing potentials in the Fayans EDF.
The neutron and proton central potentials are given by
\begin{align}
U_t(\mathbf{r}) &= \frac{\delta{\cal E}[\rho,\mathbf{J},\tilde{\rho},\tilde{\rho}^\ast]}{\delta \rho_t}\nonumber \\
&=
\frac{\delta{\cal E}_{\rm v}^{\rm Fy}[\rho]}{\delta\rho_t} +
\frac{\delta{\cal E}_{\rm s}^{\rm Fy}[\rho,(\nabla\rho)^2]}{\delta\rho_t} +
\frac{\delta{\cal E}_{\rm ls}^{\rm Fy}[\rho,\mathbf{J}]}{\delta\rho_t} 
\nonumber \\ &\quad + 
\frac{\delta{\cal E}_{\rm Coul}[\rho]}{\delta\rho_t} +
\frac{\delta{\cal E}_{\rm pair}^{\rm Fy}[\rho,\tilde{\rho},\tilde{\rho}^\ast]}{\delta\rho_t},
\label{eq:central}
\end{align}
where 
\begin{align}
\frac{\delta{\cal E}_{\rm v}^{\rm Fy}}{\delta\rho_t}
&= \frac{\epsilon_F}{3} a_+^{\rm v} \left[
- \sigma \frac{ h_{1+}^{\rm v} + h_{2+}^{\rm v}}{\{1 + h_{2+}^{\rm v} [x_0(\mathbf{r})]^\sigma \}^2} [x_0(\mathbf{r})]^{\sigma+1} \right.
\nonumber \\ &\quad 
\left.
+2 \frac{1 - h_{1+}^{\rm v} [x_0(\mathbf{r})]^\sigma}{1 + h_{2+}^{\rm v} [x_0(\mathbf{r})]^\sigma} x_0(\mathbf{r})
\right] \nonumber \\
&\quad 
+ \frac{\epsilon_F}{3}a_-^{\rm v}
\left[
-\frac{ h_{1-}^{\rm v} + h_{2-}^{\rm v}}{ [1+ h_{2-}^{\rm v} x_0(\mathbf{r})]^2} [x_1(\mathbf{r})]^2 \right.
\nonumber \\ & \quad \left. \pm 2\frac{1 - h_{1-}^{\rm v}x_0(\mathbf{r})}{1+h_{2-}^{\rm v} x_0(\mathbf{r})}
x_1(\mathbf{r})
\right],  \label{eq:Fayansvolumepot}
\\
\frac{\delta{\cal E}_{\rm s}^{\rm Fy}}{\delta\rho_t} &= 
-\frac{\epsilon_F}{3}
\frac{ \sigma h_+^{\rm s} a_+^{\rm s} r_s^2 [\nabla x_0(\mathbf{r})]^2
[x_0(\mathbf{r})]^{\sigma-1}
}{\left\{1+h_+^{\rm s} [x_0(\mathbf{r})]^\sigma+h_\nabla^{\rm s} r_s^2 [\nabla x_0(\mathbf{r})]^2\right\}^2}  \nonumber \\
&\quad -\frac{2\epsilon_F}{3\rho_{\rm sat}}
\frac{ a_+^{\rm s} r_s^2 \left\{ 1 + h_+^{\rm s} [x_0(\mathbf{r})]^\sigma\right\}\Delta \rho_0(\mathbf{r})}
{ \left\{
1 + h_+^{\rm s} [x_0(\mathbf{r})]^\sigma + h_\nabla^{\rm s} r_s^2 [\nabla x_0(\mathbf{r})]^2
\right\}^2}
\nonumber \\
& \quad +
\frac{ \epsilon_F a_+^{\rm s} r_s^2 h_+^{\rm s} \sigma[x_0(\mathbf{r})]^{\sigma-1}
[\nabla \rho_0(\mathbf{r})]^2
}{ 3\rho_{\rm sat}^2} \nonumber \\ 
& \quad\times
\frac{1 + h_+^{\rm s} [x_0(\mathbf{r})]^\sigma - 3h_\nabla^{\rm s} r_s^2 [\nabla x_0(\mathbf{r})]^2
}{
\left\{
 1 + h_+^{\rm s} [x_0(\mathbf{r})]^\sigma + h_\nabla^{\rm s} r_s^2 
 [\nabla x_0(\mathbf{r})]^2 \right\}^3
}
\nonumber\\
&\quad + \frac{8\epsilon_F}{3\rho_{\rm sat}^3}
 \frac{ a_+^{\rm s} h_\nabla^{\rm s} r_s^4 \left\{ 1 + h_+^{\rm s} [x_0(\mathbf{r})]^\sigma\right\}}
{\left\{
 1 + h_+^{\rm s} [x_0(\mathbf{r})]^\sigma + h_\nabla^{\rm s} r_s^2 [\nabla x_0(\mathbf{r})]^2
\right\}^3} \nonumber \\ 
& \quad\times
\sum_{ab} [\nabla_a \rho_0(\mathbf{r})][\nabla_b \rho_0(\mathbf{r})][\nabla_a\nabla_b \rho_0(\mathbf{r})], \\
\frac{\delta{\cal E}_{\rm ls}^{\rm Fy}}{\delta \rho_t} 
&= \frac{4\epsilon_F r_s^2}{3\rho_{\rm sat}} [\kappa \mathbf{\nabla}\cdot\mathbf{J}_0(\mathbf{r})  \pm \kappa'\mathbf{\nabla}\cdot\mathbf{J}_1(\mathbf{r})],  \label{eq:Fayanslspot} \\
    \frac{\delta{\cal E}_{\rm Coul}}{\delta\rho_{\mathrm{p}}}
    &= \frac{e^2}{2} \int d^3 r' \frac{\rho_{\mathrm{p}}(\mathbf{r}')}{|\mathbf{r}-\mathbf{r}'|} \nonumber \\ &\quad
   -\left(\frac{3}{\pi}\right)^{\frac{1}{3}} e^2
    [\rho_{\mathrm{p}}(\mathbf{r})]^{\frac{1}{3}}
    \left\{1 - h_{\rm Coul} [x_0(\mathbf{r})]^\sigma\right\} \nonumber \\ &\quad 
    + \sigma\frac{h_{\rm Coul}}{\rho_{\rm sat}}
    \frac{3}{4}\left(\frac{3}{\pi}\right)^{\frac{1}{3}} e^2 [\rho_{\mathrm{p}}(\mathbf{r})]^{\frac{4}{3}} [x_0(\mathbf{r})]^{\sigma-1}, \\
    \frac{\delta{\cal E}_{\rm Coul}}{\delta\rho_{\mathrm{n}}}
    &= \sigma\frac{h_{\rm Coul}}{\rho_{\rm sat}}
    \frac{3}{4}\left(\frac{3}{\pi}\right)^{\frac{1}{3}} e^2 [\rho_{\mathrm{p}}(\mathbf{r})]^{\frac{4}{3}} [x_0(\mathbf{r})]^{\sigma-1}.
\end{align}
The last term in Eq.~(\ref{eq:central}) is the pairing rearrangement term given in Eq.~(\ref{eq:rearrangement}).
The spin-orbit potential is given by
\begin{align}
W_t(\mathbf{r}) = \frac{\delta{\cal E}_{\rm ls}^{\rm Fy}}{\delta(\mathbf{\nabla}\cdot\mathbf{J}_t)}
= \frac{4\epsilon_F r_s^2}{3\rho_{\rm sat}} [ \kappa \rho_0(\mathbf{r}) \pm \kappa' \rho_1(\mathbf{r})]. 
\label{eq:Fayanslspot2}
\end{align}
The plus (minus) sign is taken for the neutron (proton) potential
in Eqs.~(\ref{eq:Fayansvolumepot}), (\ref{eq:Fayanslspot}), and (\ref{eq:Fayanslspot2}).

The pairing potential is given by
\begin{align}
  \tilde{U}_t(\mathbf{r}) &=  \frac{\delta{\cal E}_{\rm pair}^{\rm Fy}[\rho,\tilde{\rho},\tilde{\rho}^\ast]}{\delta\tilde{\rho}^\ast_t} \nonumber \\ &
  = \frac{4\epsilon_F}{3\rho_{\rm sat}}\left[
    f_{\rm ex}^\xi + h_+^\xi [x_{\rm pair}(\mathbf{r})]^\gamma + h_\nabla^\xi r_s^2 [{\nabla}x_{\rm pair}(\mathbf{r})]^2
  \right] 
  \nonumber \\ & \quad \times \tilde{\rho}_t(\mathbf{r}).
\end{align}

\section{Charge density\label{sec:chargedensity}}

The charge density is computed from the convoluted single-nucleon densities~\cite{Kurasawa2019},
\begin{align}
\rho_{\rm ch}(\mathbf{r}) = \sum_{t={\mathrm{n},\mathrm{p}}}
[ \rho_{ct}(\mathbf{r}) + \omega_{ct}(\mathbf{r})],
\end{align}
where 
\begin{align}
    \rho_{ct}(\mathbf{r}) = \int d^3 x \int \frac{d^3q}{(2\pi)^3}
    e^{i\mathbf{q}\cdot(\mathbf{x}-\mathbf{r})} G_{Et}(q^2) \rho_t(\mathbf{x}),
    \label{convolution}\\
    \omega_{ct}(\mathbf{r}) = \int d^3 x \int \frac{d^3 q }{(2\pi)^3}
    e^{i\mathbf{q}\cdot(\mathbf{x}-\mathbf{r})} F_{2t}(q^2) \omega_t(\mathbf{x}).
\end{align}
In Eq.~\eqref{convolution}, $\rho_t(\mathbf{r})$ is the point proton or neutron density.
We use the non-relativistic form for the spin-orbit density,
\begin{align}
    \omega_t(\mathbf{r}) = - \frac{ \mu'_t \hbar^2}{2 (Mc)^2}
    \mathbf{\nabla}\cdot\mathbf{J}_t(\mathbf{r}),
\end{align}
where $Mc^2=939$ MeV is the averaged nucleon mass,
and $\mu'_t$ is the anomalous magnetic moment,
related to the nucleon magnetic moments $\mu_{\mathrm{n}} = -1.913$ and $\mu_{\mathrm{p}} = 2.793$ as $\mu'_{\mathrm{n}} = \mu_{\mathrm{n}}$ and $\mu'_{\mathrm{p}} = \mu_{\mathrm{p}}-1$.
The Sachs and Pauli form factors are chosen to be
\begin{align}
    G_{E\mathrm{n}}(q^2) &= \frac{1}{(1 + r_+^2 q^2/12)^2} 
    - \frac{1}{(1 + r_-^2q^2/12)^2}, \\
    G_{E\mathrm{p}}(q^2) &= \frac{1}{(1 + r_{\mathrm{p}}^2 q^2/12)^2}, \\
    F_{2\mathrm{n}}(q^2) &= \frac{ G_{E\mathrm{p}}(q^2) - G_{E\mathrm{n}}(q^2)/\mu'_{\mathrm{n}}}{1 + q^2/4M^2},\\
    F_{2\mathrm{p}}(q^2) &= \frac{ G_{E\mathrm{p}}(q^2)}{1 + q^2/ 4M^2},
\end{align}
and $r_{\mathrm{p}}=0.84$ fm \cite{Workman2022}, $r_{\pm}^2 = (0.830)^2 \mp 0.058$ fm$^2$ \cite{Kurasawa2021} are used.

The charge radius is computed from the charge density distribution,
\begin{align}
  \langle r^2\rangle_{\rm ch} &= \frac{1}{Z}\int d^3r\,r^2 \rho_{\rm ch}(\mathbf{r})
  \\
  &= \langle r^2\rangle_{\rm p} + r_{\rm p}^2 + \frac{N}{Z}(r_+^2 - r_-^2)
  + \delta\langle r^2\rangle_{c},
  \label{charge.radius}
\end{align}
where
\begin{align}
  \langle r^2\rangle_{\rm p} &= \frac{1}{Z}\int d^3r\,r^2 \rho_{\rm p}(\mathbf{r}),
\end{align}
and the relativistic correction $\delta\langle r^2\rangle_c$ is,
\begin{align}
\delta\langle r^2\rangle_c &= \frac{3\hbar^2}{4(Mc)^2}
+ \big(1+\frac{1}{2\mu'_{\mathrm{p}}}\big) \frac{1}{Z} \int d^3r\,r^2 \omega_{\mathrm{p}}(\mathbf{r}) \nonumber\\
&\quad + \frac{1}{Z}\int d^3r\,r^2 \omega_{\mathrm{n}}(\mathbf{r}).
\end{align}
Though not shown here, the integral of $r^2 \omega_t(\mathbf{r})$ can be expressed by the expectation value of $\boldsymbol{\ell}\cdot\hat{\bm{\sigma}}$
($\boldsymbol{\ell}=\mathbf{r}\times\mathbf{p}$).

\bibliographystyle{plain}
\bibliography{Ca_charge}

\end{document}